\documentclass[prb,twocolumn,10,aps,superscriptaddress,longbibliography]{revtex4-1}
\usepackage[T1]{fontenc}
\usepackage[latin9]{inputenc}
\usepackage{color}
\usepackage{float}
\usepackage{amsmath}
\usepackage{amssymb}
\usepackage{graphicx}
\usepackage{esint}
\usepackage{epstopdf}
\usepackage[colorlinks,bookmarks=false,citecolor=blue,linkcolor=blue,urlcolor=blue,hyperfootnotes=true]{hyperref}

\makeatletter
\pdfoutput = 1

\usepackage{amsfonts}
\usepackage[normalem]{ulem}
\newcommand{\bq}{\mathbf{q}}

\newcommand{\pf}{2{\bf {p}}_{\text{F}}}

\newcommand{\bbq}{\mathbf{Q}}
\newcommand{\kv}{{\bf {k}}}

\newcommand{\pff}{2\mathbf{\bold{p_{\text{F}}}}(\bold{k})}
\newcommand{\pffq}{2\mathbf{\bold{p_{\text{F}}}}(\bold{k+q})}
\newcommand{\pfff}{2\mathbf{\bold{p_{\text{F}}}}(-\bold{k})}

\makeatother

\begin{document}

\title{A model for the neutron resonance in  HgBa$_{2}$CuO$_{4+\delta}$}

\date{\today}

\author{X.\ Montiel}
\email{xavier.montiel@rhul.ac.uk}
\affiliation{Institut de Physique Th\'{e}orique, L'Orme des Merisiers, CEA-Saclay, 91191 Gif-sur-Yvette, France }
\affiliation{Department of Physics, Royal Holloway, University of London, Egham, Surrey TW20 0EX, United Kingdom}

\author{C.P\'{e}pin}
\email{catherine.pepin@cea.fr}
\affiliation{Institut de Physique Th\'{e}orique, L'Orme des Merisiers, CEA-Saclay, 91191 Gif-sur-Yvette, France }
\begin{abstract}
We study the spin dynamics of the Resonant Excitonic State (RES) proposed,
within the theory of an emergent SU(2) symmetry, to explain some properties
of the pseudo-gap phase of cuprate superconductors. The RES can be
described as a proliferation of particle-hole patches with an internal
modulated structure. We model the RES modes as a charge order with
multiple $\pf$ ordering vectors, where $\pf$ connects two opposite
side of the Fermi surface. This simple modelization enables us to
propose a comprehensive study of the collective mode observed at the
antiferromagnetic (AF) wave vector $\bbq=(\pi,\pi)$ by Inelastic
Neutron Scattering (INS) in both superconducting state (SC), but also
in the Pseudo-Gap regime. In this
regime, we show that the dynamic spin susceptibility accuses a loss
of coherence terms except at special wave vectors commensurate with
the lattice. We argue that this phenomenon could explain the change
of the spin response shape around $\bbq$. We demonstrate that the
hole dependence of the RES spin dynamics is in agreement with the
experimental data in HgBa$_{2}$CuO$_{4+\delta}$.  
\end{abstract}
\maketitle

\section{Introduction}

\begin{figure}[!h]
\includegraphics[width=7cm]{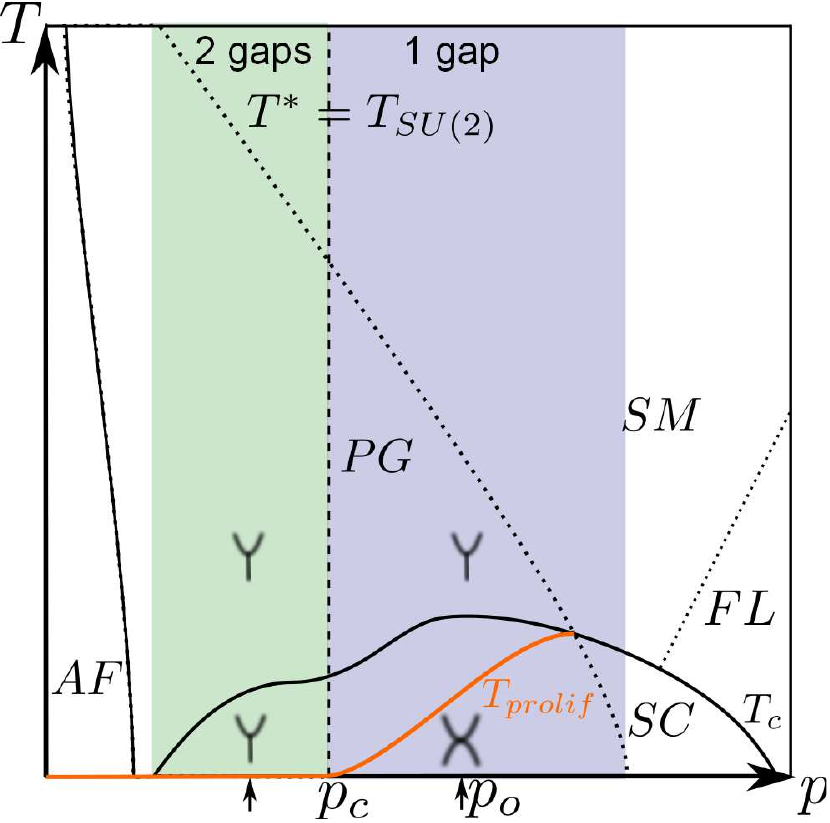}
\caption{\label{phased}(Color online) \textcolor{black}{Schematic Temperature-
doping (T,p) phase diagram of hole-doped cuprate compounds. The antiferromagnetic
(AF) phase develops close to half filling ($p=0$). The SC phase appears
at intermediate hole doping and $T_{c}$ is maximum at optimal doping
$p_{o}$. Above $T_{c}$, in the underdoped regime, $p<p_{o}$, the
system exhibits a large pseudo-gap (PG) phase until the temperature
$T^{*}=T_{SU(2)}$. The RES model explains the PG phase by the proliferation
of local excitonic patches above the temperature $T_{prolif}$ (orange
line) \cite{Montiel_1611}. The proliferation temperature vanishes below
the doping $p_{c}$. Below $p_{c}$, the preformed particle hole pairs
in the RES are more stable than the Cooper pairs. Consequently, the
Fermi surface is fractionalized : the Cooper pairs develop in the
N region only while the preformed particle-hole pairs populate in the
AN region. The system exhibits a two-gap regime (green area). For
doping above $p_{c}$, the particle-hole pairs are less stable than
the Cooper pairs and the SC gaps out the whole Fermi surface. The
system exhibits a one-gap regime (blue area).The two black arrows
represent the two hole doping where further calculations are performed.
left arrow) Below $p_{c}$, we have $T_{prolif}=0$ then RES is strong
compared to SC and the AN region is massively gapped by RES mechanism
and $\Delta_{RES}>\Delta_{SC}$ in the AN region of the first BZ.
It is a two-gaps regime and the energy fluctuation spectrum of the
spin susceptibility exhibits the same Y-shape in both the PG and the
SC phase. right arrow) Close to optimal doping (for $p_{c}<p$), we
have $T_{prolif}\protect\neq0$. The RES is weaker than SC state and
$\Delta_{RES}<\Delta_{SC}$ in the AN region of the first BZ. Consequently
in the SC state, the AN zone is nearly completely gapped out by Cooper
pairs. It is a one-gap regime where the energy fluctuation spectrum
exhibit a X-shape in the SC phase which transforms itself in a Y-shape
above $T_{c}$.}}
\end{figure}

Inelastic Neutron Scattering (INS) and Electronic Raman spectroscopy (ERS)
are experimental probes based on two particles processes which allow
the observation of coherence effects, like the superconducting (SC)
coherence peak whose energy is proportional to the transition temperature
$T_{c}$, or the emergence of collective modes, which act a signature
of the symmetries of the system. The study of collective modes could
be a key to reveal the physical mechanisms at the origin of high critical
temperature SC of cuprate compounds. A long standing mystery of such
compounds is the pseudo-gap (PG) phase which exists in the underdoped
regime \cite{Alloul89,Warren89,Tallon01} (see Fig. \ref{phased})
and manifests itself by a loss of electronic density of states, without
being related to any obvious symmetry breaking.

The presence of a collective spin resonance around the antiferromagnetic
(AF) wave vector in the SC state has been
known long ago. It has first been observed by INS experiments around
the AF wave vector ${\bf {Q}}$$=(\pi,\pi)$ at a frequency $\omega_{res}=41\,meV$
in YBa$_{2}$Cu$_{3}$O$_{7-\delta}$ (YBCO) \cite{Keimer89,Keimer96,Rossat,Mook93,Fong95,Bourges19052000,Sidis2004,Hinkov04,Dai99,Dai97,Fong97,Fong00}
and at similar energies in other compounds \cite{Hayden89,Bourges2005,Sidis2001,Hayden96,Hayden93}
in the superconducting (SC) phase.\textcolor{black}{{} In this paper,
we focus our study on recent experiments performed in monolayer Hg-based
cuprate compounds: HgBa$_{2}$CuO$_{4+\delta}$ (Hg-1201) \cite{Greven2016}.
This compound represents a perfect playground to study the physics
of cuprates superconductors. It is a single CuO$_{2}$ layer that
allows to neglect the effect of interlayer coupling of multilayered
systems as well as the effect  charge reservoir, such as CuO chains in YBCO or incommensurate BiO layer in Bi$_{2}$Sr$_{2}$CaCu$_{2}$O$_{o+\delta}$ (Bi-2212) compounds. It exhibits the universal resonance around the AF wave vector $\bbq$, which shows three main
features.}

1) The resonance stands below the $2\Delta_{SC}^{0}$
threshold of particle-hole continuum ($\Delta_{SC}^{0}$ is the maximum
of the $d$-wave superconducting (SC) gap) and the frequency resonance $\omega_{res}$ decreases with underdoping \cite{Greven2016}. Moreover,
a precursor of this resonance exists in the PG above $T_{c}$, the
SC critical temperature \cite{Greven2016} where the resonance is
observed at the same frequency $\omega_{res}$ than in the SC state
with a lower intensity. The latter has also been observed in other
cuprate compounds \cite{Dai99,Dai97,Fong97,Fong00,Stock04,Stock05,Hayden04}.

2) The energy fluctuation spectrum around $\bbq$
has a peculiar behavior and distribution  in phase space in the underdoped regime \cite{Greven2016}.
The low and high energy parts of the fluctuation spectrum behave differently
with temperature. The high energy part (for $\omega\gtrsim\omega_{res}$)
of the energy fluctuation spectrum does not change across $T_{c}$
or $T^{*}$, the pseudo-gap (PG) critical temperature. This behavior
most probably corresponds to the response of localized spins which
originate the proximity of the AF phase. On the other hand, the low
energy part ($\omega\lesssim\omega_{res}$) of the energy fluctuation
spectrum changes across $T_{c}$. Below, $T_{c}$, a gap opens around
$\bbq$ and the intensity of the resonance increases from $T_{c}$ and $T=0$. Moreover, two branches appear from either side of the momentum $\bbq$ and meet in $\bbq$ at $\omega=\omega_{res}$ forming the so-called X-shape- also called ``hourglass''-shape. Above $T_{c}$,
the gap at $\bbq$ closes and the two energy branches disappear, forming
the so called Y-shape- while the intensity of the resonance decreases
until $T^{*}$. This feature has been observed in other cuprate compounds
\cite{Hayden04,Tranquada04,Vignolle07,Hinkov07,Tranquada09}.

3) A very specific doping dependence of the spin
fluctuations is reported in monolayer Hg-based cuprate compound Hg-1201
\cite{Greven2016}. In the underdoped regime, at hole doping below
$0.12$ ($p<0.12$), a Y-shape has been observed close to the vector
$\bbq$ in both the PG and the SC phase  without any change at $T_{c}$. For higher doping, $p\ge0.12$,
the X-shape is recovered in the SC phase. A summary of the different features is presented in Fig.\ref{phased}

Several models have been proposed to explain this
collective mode \cite{Onufrieva:2000tk,Onufrieva:2002uf,Brinckmann99,Demler95,Norman07,NormanChub01,Pepin94,Eremin01,Eremin:2005ba,Eschrig:2000bf,Eschrig:2002wy,Eschrig:2003aa}.
An exhaustive review of all these approaches is presented in Ref.\cite{Eschrig_rev}. Among various scenarii to account for the spin excitation spectrum in the SC state,  the INS resonance was ascribed to SO(5) emergent symmetry as a $\pi$-collective
mode \cite{Demler95,Demler04} relating SC to AF order. However, it
has been shown that the $\pi$-mode has an anti-bounding with the optical
mode which pushes it at a higher energy than experimentally observed
\cite{NormanChub01}. The most commonly accepted explanation within the framework of itinerant magnetism, is that
the INS resonance is a particle-hole bound state below the spin
gap (a spin-triplet exciton) which is stabilized  by repulsive interaction left within the $d$-wave SC state. \cite{NormanChub01,Pepin94,Eschrig:2000bf,Eremin01,Chubukov01aa,Eschrig:2002wy,Abanov02,Eschrig:2003aa}.
This scenario well reproduces the structure of the spin excitation
in the SC state in the optimally and overdoped regime. In the underdoped
regime, the observation of the INS resonance in the PG state above
$T_{c}$ leads to a more complex situation. The shape of the resonance
changes from ``X'' to ``Y'' with the presence of some additional
spectral weight in the vicinity of $\bbq$, whereas, in Hg-1201, the energy of
the collective mode remains unchanged compared to the SC phase (see Fig. \ref{phased}). This
observation is very difficult to account for theoretically.Recently, an incommensurate spiral spin order stabilized by quantum fluctuations upon doping the AF Mott insulator has been proposed to explain the evolution of the energy fluctuation spectrum around $\bbq$ with doping in YBCO \cite{Onufrieva2017}. The main
difficulties lies on a correct modelization of the PG phase which,-if
we believe the excitonic explanation in the SC phase, has to retain
a certain amount of coherence if the collective mode is to be observed
at all in this regime.

In parallel, ERS measurments in Hg-1201 provides very interesting
and complementary information for the study of collective modes in
the underdoped regime. A noticeable change of behavior is observed
in Raman data around $0.12$ hole doping. Raman scattering is a dynamical
response, which probes the charge channel at $q=0$. Moreover, specific
structure factors enable to scan the Brillouin zone with respect to
respective symmetries : the $A_{1g}$ response is isotropic, the $B_{1g}$
symmetry scans the anti-nodal (AN) regions $\left(0,\pm\pi\right)$
and $\left(\pm\pi,0\right)$, while the $B_{2g}$ symmetry selects
the nodal (N) region $\left(\pm\pi/2,\pm\pi/2\right)$ \cite{Devereaux-RMP}.
For doping $p<0.12$, the Raman data exhibits a large SC coherence
peak in the $B_{2g}$ symmetry, while its intensity is very low in
the $B_{1g}$ symmetry. For higher doping, $p\ge0.12$, the SC coherent
peak has a huge intensity in the $B_{1g}$ symmetry and decreases
in the $B_{2g}$ symmetry \cite{LeTacon108,LeTacon111}. This change
of behavior around the same doping in both Raman and INS probes suggests
that the coherence effect that are getting lost around $T_{c}$ are
a key in the explanation of the feature 3). To the best of our knowledge,
the feature 3) has only been observed in Hg-1201 compound.

Here, we calculate the two-particle responses in both charge and spin sectors and compare them with experimental observations reported by ERS and INS 
in the underdoped regime, within a new theoretical explanation for
the of the PG phase : the Resonant Excitonic State (RES) which can
be described as preformed excitonic (particle-hole) pairs \cite{Kloss15a,Montiel_1611}. Although different theoretical approaches have been developed to explain
the PG phase, as stated above, the issue of the change of shape of
the INS resonance across $T_{c}$ has never been addressed before
and a comprehensive study of the relations between neutron and Raman
susceptibilities in this region are given here for the first time.
There have been many proposals for the PG phase of the cuprates, based
on AF fluctuations \cite{abanov03,Norman03,sfbook}, strong correlations
\cite{Lee06,Gull:2013hh,Sorella02}, loop current \cite{Kotliar90,Varma97}
or emergent symmetry models \cite{Zhang97,Demler04}. A recent study proposes explain the PG phase with a SU(2) emergent symmetry model
where the SU(2) symmetry relates the SC state to the charge sector
\cite{Metlitski10b,Efetov13}. The PG phase is then described by a
composite $d$-wave SC and charge order parameter and the SU(2) symmetry
is restored by thermal fluctuations \cite{Metlitski10b,Efetov13}.

Recent investigations demonstrated that SU(2) symmetry could emerge from short-range AF interactions \cite{Kloss15a,Montiel_1611}.
Proceeding by integrating out the SU(2) pairing fluctuations,
we describe the PG state as a new type of charge order called Resonant
Excitonic State (RES) \cite{Kloss15a,Montiel_1611}. The RES can be described as
excitonic (particle-hole pair) patches with an internal checkerboard
charge modulation. In this scenario, the PG originates SU(2) pairing
fluctuations and the whole physics in the underdoped regime is governed
by SU(2) symmetry. Such a scenario naturally associates the resonance
observed in the PG phase with the underlying SU(2) symmetry. The novelty
of our approach is that a certain form of coherence is retained in
the PG phase, at specific wave vector commensurate with the lattice.
The role of the underlying SU(2) symmetry is essential here in two ways.
It preferentially selects a $d$-wave form factor for the pseudo-gap,
thus allowing a change of sign between anti-nodal $\left(0,\pm\pi\right)$
and $\left(\pm\pi,0\right)$ regions related by the vector $\bbq$.
It also restricts the gapping out of the Fermi surface to a small
region around the anti-nodes, which leads to the emergence of spectral
weight around $\bbq$. The specific ``Y'' shape of the resonance
in the PG phase, with the elongated tail at $\bbq$ is specific, within
our theory, of the particle-hole excitons with many $\pf$ wave vectors.

This paper is divided as follow : in the section
\ref{model}, we present the theoretical model we have used to model
the RES and the SC state and we explain how we calculate the spin
susceptibility and the Raman susceptibility. In section \ref{results},
we present our results. In the section \ref{discussion}, we present
a discussion of our results and a comparison with the experimental
data before to conclude in the section \ref{conclusion}.

\section{The Theoretical model \label{model} }

In this section, we present the minimal model that describes the Resonant
Excitonic State (RES). The RES is a recent scenario proposed to explain
the PG phase\cite{Kloss15a,Montiel_1611}.

In this model, we modelize the RES as a charge ordering state
with multiple $\pf$ ordering vectors. The $\pf$ vectors connect two opposite
sides of the Fermi surface (see Fig. \ref{FS}). The $\pf$ vectors depend on the momentum $\bf{k}$ in the first BZ and write as a function of momentum $\bf{k}$: $\pff$. In the following, we assume that the $\pf$ vector of a point far from the Fermi surface is the $\pf$
vector of the closest point of the Fermi surface. Note that on the Fermi surface, we have $\pf({\bf{k}_{F}})=-2\bf{k}_{F}$ which implies that $\bf{k}_{F}-\pf(\bf{k}_{F})=-\bf{k}_{F}$.  The vectors $\pff$ are represented in Fig. \ref{FS}. The $\pf$ structure
corresponds to charge modulations with multiple wave vectors, which
creates local modes, also called ``patches'' or ``droplets'' of
particle-hole pairs. 

Here we study how the proliferation of those
modes can account, phenomenologically, for the INS spectrum around
${\bf {Q}}$ in the SC and in the PG phases, where the PG phase is
described by a RES, in competition with the SC phase. At low temperature,
SC and RES coexist until $T_{c}$, forming a kind of super solid.
Moreover the proliferation temperature for the local RES modes is
doping dependent, as depicted as the orange line in Fig.\ref{phased}.
For $p<0.12$, the proliferation temperature is
extremely small, meaning that the whole AN region of the BZ is dominated
by the RES, and the superconductivity comes mainly from the region
around the nodes. For 0.12<p<0.25 the proliferation temperature is
non zero, which means that at low temperature we are inside a ``one
gap'' SC phase. Above $T_{c}$ until $T^{*}$ only the RES remains
\cite{Kloss15a,Montiel_1611}.

An important point, in our scenario, is that the binding force leading
to the formation of the particle-hole pairs, is the SU(2) fluctuations
between the SC state and the charge sectors. While these fluctuations
are gapped in the SC phase, they become important in the PG phase
and in the underdoped region, which leads to the formation, and proliferation
of patches - or droplets- of excitonic particle-hole pairs above a
characteristic temperature, here $T_{prolif}$.

\subsection{The RES minimal model }
\label{min} 
The simplified version of our theory consists of performing
a mean field decoupling of a Hamiltonian retaining short range AF
correlations in the charge and SC channels.
\begin{align}
{\cal H}_{start} & =\sum_{i,j,\sigma}t_{ij}c_{i\sigma}^{\dagger}c_{j\sigma}+\frac{1}{2}\sum_{\left\langle i,j\right\rangle }J_{ij}\mathbf{S}_{i}.\cdot\mathbf{S}_{j},\label{eq:1}
\end{align}

where $t_{ij}$ is the hopping from site $i$ to $j$ and $J_{ij}$ the AF super-exchange
between spins $\mathbf{S}_{i}=\sum_{\alpha\beta}c_{i\alpha}^{\dagger}\vec{\sigma}_{\alpha\beta}c_{i\beta}$ sitting
on nearest neighbors $\left\langle i,j\right\rangle$ \cite{Brinckmann99,Brinckmann01}. $\vec{\sigma}_{\alpha\beta}$ is a set of Pauli Matrices and $c^{(\dagger)}_{i,\sigma}$ is the annihilation (creation) operator of an electron  with the spin $\sigma$ on site $i$. Note that the effects of strong on-site Coulomb repulsion and thus the double occupancy, are so far neglected in this model.
Applying the Fourier transform on the fermionic operator $c_{i,\sigma}=\frac{1}{\sqrt{N}}\sum_{\bf{k}}c_{\bf{k},\sigma}e^{i\bf{k.r_{i}}}$, the Hamiltonian in Eq. (\ref{eq:1}) becomes :
\begin{align}
\begin{array}{c}
{\cal H}_{start}  =\sum_{\bf{k},\sigma}\xi_{\bf{k}}c_{\bf{k}\sigma}^{\dagger}c_{\bf{k}\sigma}\\
+\frac{1}{2}\sum_{\bf{k},\bf{k}',\bf{q}}J_{\bf{q}}c_{\bf{k}\alpha}^{\dagger}\vec{\sigma}_{\alpha\beta}c_{\bf{k+q}\beta}c_{\bf{k'+q}\gamma}^{\dagger}\vec{\sigma}_{\gamma\delta}c_{\bf{k'}\delta}.\label{eq:1fourier}
\end{array}
\end{align}
We describe the RES and SC state by the effective
action $S_{\text{eff}}=-\sum_{\mathbf{k,\sigma}}\Psi_{{\bf {k}}}^{\dagger}\hat{G}^{-1}\Psi_{{\bf {k}}}$,
in the basis $\Psi_{{\bf {k}}}^{\dagger}=\left(c_{{\bf {k},\sigma}}^{\dagger},c_{{\bf {-k}+\pfff,\overline{\sigma}}},c_{{\bf {k}+\pff,\sigma}}^{\dagger},c_{-{\bf {k},\overline{\sigma}}}\right)$
, and with
\begin{widetext}
\begin{align}
\hat{G}^{-1}\left({\bf {k},\epsilon}\right)=\left(\begin{array}{cccc}
i\epsilon-\xi_{{\bf {k}}} & 0 & \Delta_{RES,{\bf {k}}} & \Delta_{SC,{\bf {k}}}\\
0 & i\epsilon+\xi_{{\bf {-k}+\pfff}} & \Delta_{SC,{\bf {k}+\pff}}^{\dagger} & -\Delta_{RES,{\bf {k}}}\\
\Delta_{RES,{\bf {k}}}^{\dagger} & \Delta_{SC,{\bf {k}+\pff}} & i\epsilon-\xi_{{\bf {k}+\pff}} & 0\\
\Delta_{SC,{\bf {k}}}^{\dagger} & -\Delta_{RES,{\bf {k}}}^{\dagger} & 0 & i\epsilon+\xi_{-{\bf {k}}}
\end{array}\right).\label{eq:greenMat4x4}
\end{align}

\end{widetext}

\begin{figure}[h]
\begin{minipage}[c]{7cm}%
a)\includegraphics[width=6cm]{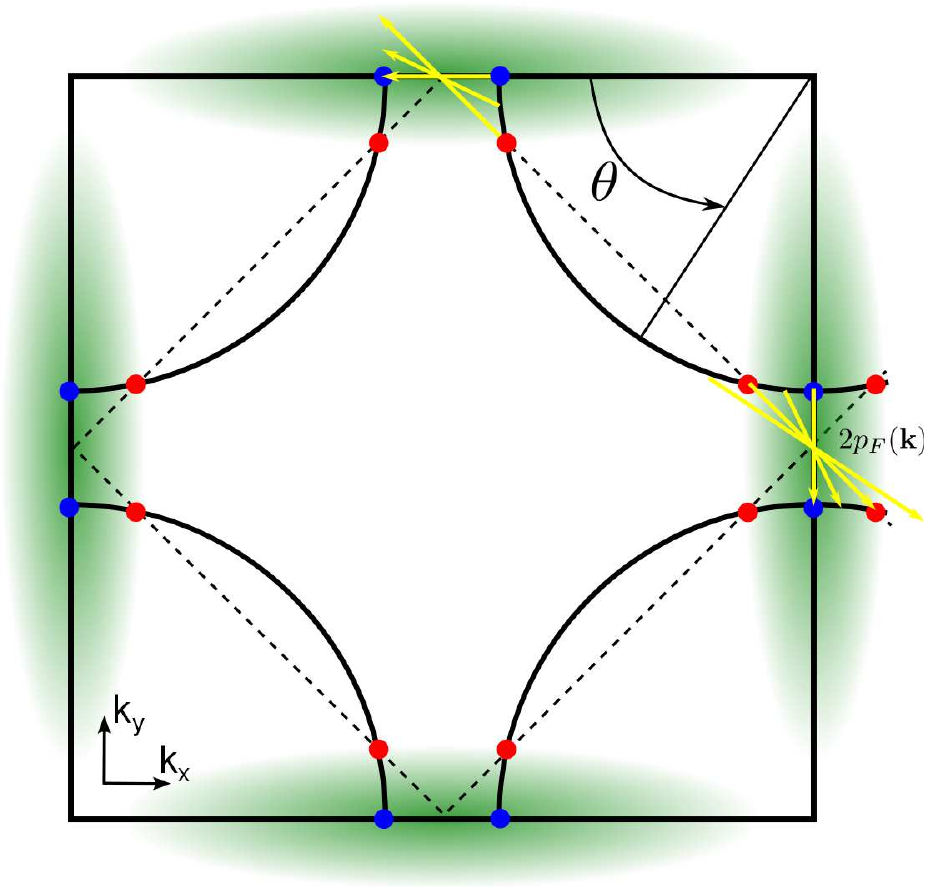}.
\end{minipage}
\begin{minipage}[c]{4.25cm}%
b)\includegraphics[width=3.95cm]{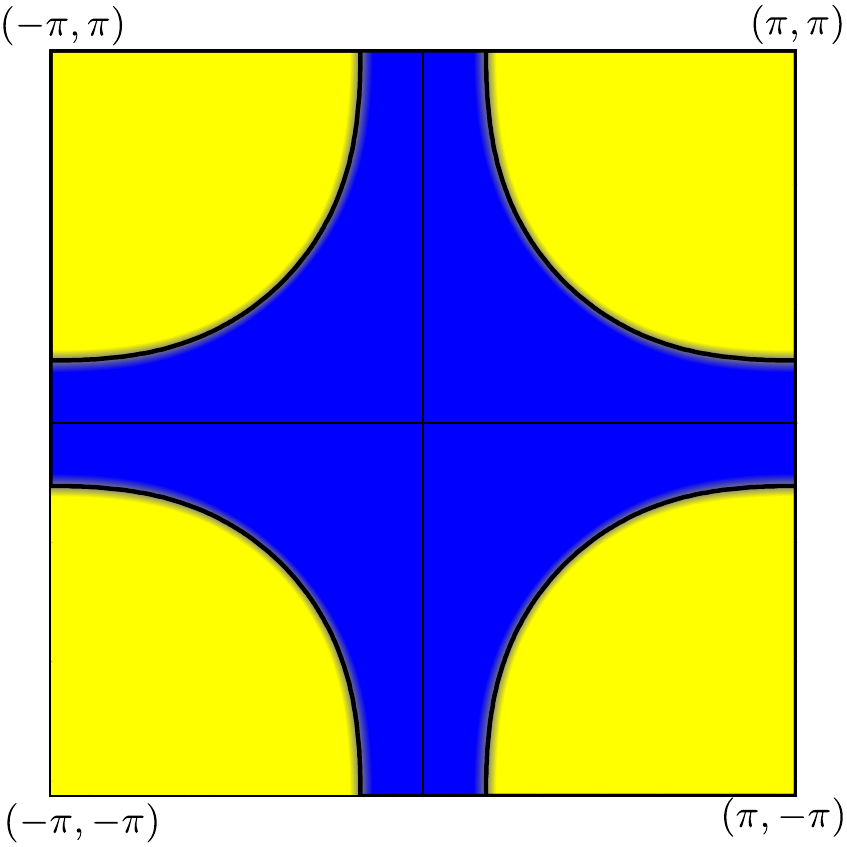}
\end{minipage}
\begin{minipage}[c]{4.25cm}%
c)\includegraphics[width=3.95cm]{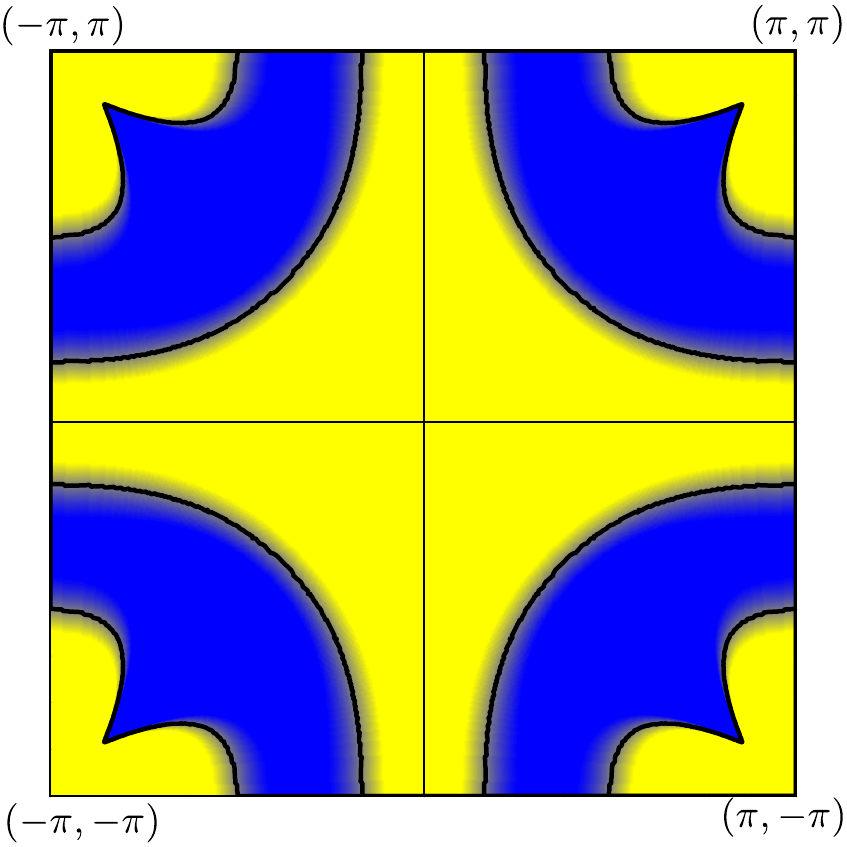}
\end{minipage}
\caption{\label{FS}(Color online) a) Schematic representation
of the hole-doped Fermi surface (solid line) in first Brillouin zone
of the square lattice. The angle $\theta$ localizes the point on
the FS. The values $\theta=0$ and $\pi/2$ represent the AN-zone
while $\theta=\pi/4$ stands for the N zone of the first BZ. The magnetic
BZ is presented in dashed line while its intersections with the Fermi
surface (the hot-spots) are the red points. The yellow arrows represent
the $\pf$ ordering vectors in the Anti-Nodal (AN) region. Note $\pf$
vector depends on the momentum ${\bf {k}}$. The RES gap develops
in the AN zone (green area). The point of the FS at the zone edge
are drawn in blue circles. In b) we represent the bare electronic
dispersion $\xi_{\mathbf{k}}$ at optimal doping $p=0.16$ while in b) we represent the $\pf$-shifted electronic dispersion $\xi_{\mathbf{k+2p_{F}(k)}}$.
The blue (yellow) area shows the electron (hole) states that have
negative (positive) energy separated by the Fermi surface (solid line).
Note that the FS of the bare and the $\pf$-shifted electronic dispersions
are the same whereas the curvature close to the FS is reversed.}
\end{figure}

Here $\xi_{{\bf {k}}}$ is the electron dispersion, including the
chemical potential $\mu$ with $\xi_{-{\bf {k}}}=\xi_{{\bf {k}}}$
and $\epsilon$ is the fermionic Matsubara frequency. $\Delta_{SC,{\bf {k}}}$
is the superconducting order parameter and $\Delta_{RES,{\bf {k}}}$
the RES one, which couples $\mathbf{k}\rightarrow \mathbf{k-\pff}$.
The RES order parameters descibes the particle-hole pairs patches that break locally the translational symmetry \cite{Kloss15a}.
 We use a tight-binding description of the electronic spectrum of Hg-1201
with $\xi_{\mathbf{k}}=-2t_{1}(\cos(k_{x}a)+\cos(k_{y}a))+2t_{2}\cos(k_{x}a)\cos(k_{y}a)+t_{3}(\cos(2k_{x}a)+\cos(2k_{y}a))+t_{4}(\cos(2k_{x}a)\cos(k_{y}a)+\cos(k_{x}a)\cos(2k_{y}a))-\mu$
where $t_{i}$ are the i$^{th}$ neighbor hopping parameters. We have
$t_{1}=-0.408\,eV$, $t_{2}=0.093\,eV$, $t_{3}=0.071\,eV$ and $t_{4}=0.036\,eV$
 (deduced from ab-initio calculations \cite{Das12}) which gives a bandwidth of $1.5\,eV$. $a$ is the elementary
cell parameter set to unity $a=1$ and $\mu$ is the chemical potential
adjusted to determine the hole doping. The Fermi surfaces of the spectrum
$\xi_{\mathbf{k}}$ and $\xi_{\mathbf{k+\pff}}$ are presented on
the figure \ref{FS}. Note that the bandwidth of the spectrum $\xi_{\mathbf{k}}$
is larger than the bandwidth of the spectrum $\xi_{\mathbf{k+\pff}}$.
We determined the Green's functions of the model by inverting the
matrix ($\ref{eq:greenMat4x4}$)

As highlighted in previous studies\cite{Kloss15a,Montiel_1611}, the
interplay of the SC and the RES order parameters is not trivial. For
intermediate temperature ($T_{c}<T<T^{*}$), only RES remains in the
system. The RES leads to the opening of a gap in the AN zone of the
first BZ, and the formation of Fermi arcs \cite{Kloss15a,Montiel_1611}. We consider a RES order
parameter with a $d$-wave symmetry as it is the SU(2) partner of the $d$-wave SC state :

\begin{align}
\Delta_{RES,\mathbf{k}}=\frac{\Delta_{RES}^{0}}{2}\gamma_{\mathbf{k}}e^{\left(-\frac{(k_{x}-\pi)^{2}a^{2}}{2\sigma_{x}^{2}}-\frac{(k_{y})^{2}a^{2}}{2\sigma_{y}^{2}}\right)},\label{DRES}\\
\mbox{ with }\gamma_{\mathbf{k}}=(cos(k_{x}a)-cos(k_{y}a)).\nonumber 
\end{align}
Here $\sigma_{x(y)}$ is the width of the Gaussian function in the
$k_{x}(k_{y})$ direction (see Fig. \ref{FS}). This
parametrization has been used to explain the opening of the PG and
the formation of Fermi arcs observed by Angle Resolved PhotoEmission
Spectroscopy (ARPES) \cite{Montiel16ARPES}.

Below $T_{c}$ ($T<T_{c}$), we define a $d$-wave gap envelop $C_{k}$
which will take into account the coexistence between the SC state
and the RES. The definition of the RES order parameter is the same
than above $T_{c}$ (see relation \ref{DRES}). 
In the following we assume that the gap envelop is related to the SC and the RES order parameters by the relation: 
\begin{align}
C_{\mathbf{k}}=\sqrt{\Delta_{SC,\mathbf{k}}^{2}+\Delta_{RES,\mathbf{k}}^{2}},\label{DSU2}
\end{align}
where $C_{\mathbf{k}}$ has a $d$-wave symmetry and a magnitude $C^{0}$:
$C_{\mathbf{k}}=\frac{C^{0}}{2}\gamma_{\mathbf{k}}$. Considering, the relation (\ref{DSU2}),
we can deduce the form of the SC order parameter, $\Delta_{SC,\mathbf{k}}=\sqrt{C_{\mathbf{k}}^{2}-\Delta_{RES,\mathbf{k}}^{2}}$.
The variation of the SC and RES order parameters
in the SC phase along the Fermi surface is presented in the Figs.
\ref{densitep10} and \ref{densitep16}. The RES develops solely
in the AN zone of the first BZ (see relation \ref{DRES}) while the
superconducting state can exist both in the nodal and anti-nodal zones.
This momentum dependence of the SC and the RES order parameters in
the first BZ is supported by electronic Raman Scattering experiments
in Hg-1201 \cite{LeTacon108,LeTacon111} or Bi-2212 \cite{Sacuto2015TS}
compounds as well as ARPES experiments \cite{Vishik12,Vishik14_He}.
The resolution of the self-consistency equation deriving from the minimal model is left for a forthcoming publication. In the following, we determine the value of the gap magnitude that reproduce the experimental data.

The real space picture is that the RES is formed of local objects,
patches or droplets, which compete with the global SC phase. When
temperature is raised, the entropy associated with the local object
is always winning compared to the energy of the global phase, in analogy
with the proliferation of vortices in a stiff quantum fluid \cite{Kosterlitz1973}. Hence
there exists a proliferation temperature for the patches of excitons,
which can be understood as follows. When the binding energy for the
formation of the Cooper pairs is greater than the energy for the formation
of the particle hole pairs, the proliferation occurs above a certain
temperature $T_{profif}\simeq\left(E_{CP}^{2}-E_{EP}^{2}\right)/g$,
where $E_{CP}$ and $E_{EP}$ is the mean field scale for the formation
of Cooper  and particle-hole pairs (at $\pf$), respectively and $g$
is a coupling constant coming from a simple Ginzburg Landau treatement \cite{Montiel_1611}. On the other hand, when $E_{CP}<E_{EP}$,
then the proliferation of exciton droplets starts at very low temperature,
which leads to $T_{prolif}\simeq0$. For a simple discussion, we identify
$E_{CP}\simeq T_{c}$ while $E_{EP}\simeq T_{SU(2)}$, which is the
energy scale associated to the SU(2) fluctuations in our theory \cite{Kloss15a,Montiel_1611}.
As depicted in Fig.\ref{phased}, there is a critical doping $p_{c}$,
situated in the underdoped region, below which $T_{prolif}\simeq0$,
whereas for $p>p_{c}$, $T_{prolif}\neq0$. The critical doping
$p_{c}$ is a crucial ingredient of our theory to explain the experimental
data in Hg-1201.

\subsection{The spin susceptibility}

We turn to the evaluation of the spin susceptibility
in the SC state and the RES. In the SC phase, we expect the spin-exciton process that explains the spin dynamics in the overdoped
part of the cuprate phase diagram to be strongly affected by the emergence
of RES in the underdoped part of the phase diagram. The spin operator writes $S_{\bf{q}}=\frac{1}{\sqrt{N}}\sum_{\bf{k}}c^{\dagger}_{\bf{k},-\sigma}c_{\bf{k+q},\sigma}$ which destroy a bosonic excitations at momentum $\bf{q}$ with a charge 0 and spin 1. Rewritting the Hamiltonian Eq. (\ref{eq:1fourier}) with the spin operator, we get ${\cal H}_{start}  =\sum_{\bf{k},\sigma}\xi_{\bf{k}}c_{\bf{k}\sigma}^{\dagger}c_{\bf{k}\sigma}+\frac{1}{2}\sum_{\bf{q}} J_{\bf{q}}S^{\dagger}_{\bf{q}}S_{\bf{q}}$. The spin susceptibility is derived from the linear response of the spin operator and reads $\chi_{S}=-i\theta\left(t\right)\langle S^{\dagger}_{\bf{q}}\left( t\right) S_{\bf{q}}\left(0\right)\rangle$. Within the Random Phase Approximation (RPA), the full spin susceptibility writes :
\begin{equation}
\chi_{S}(\omega,\mathbf{q})=\frac{\chi_{S}^{0}(\omega,\mathbf{q})}{1+J(\mathbf{q})\chi_{S}^{0}(\omega,\mathbf{q})}\label{eq:sus}
\end{equation}
with $J(\mathbf{q})=2J_{0}\left(\text{cos}(q_{x}a)+\text{cos}(q_{y}a)\right)$ due to
exchange between near-neighbor copper sites.
In the equation (\ref{eq:sus}), $\chi_{S}^{0}$ is the bare polarization
bubble constructed from the Green's function and $J(\mathbf{q})$
is super-exchange interaction from Eqn.(\ref{eq:1}). Note that full diagrammatic contributions to the bare susceptibility is discussed in Appendix \ref{feynman}. The bare polarization
can be evaluated by the formula \cite{Norman07,Schrieffer64}: 
\begin{equation}
\chi_{S}^{0}\left(\omega,\mathbf{q}\right)=-\frac{T}{2}\sum_{\epsilon,\mathbf{k}}\text{Tr}\left[\hat{G}\left(\omega+\varepsilon,\mathbf{k+q}\right)\hat{G}\left(\varepsilon,\mathbf{k}\right)\right]\label{eq:susgen}
\end{equation}
where $\varepsilon(\omega)$ is the fermionic (bosonic) Matsubara
frequency, $\mathbf{k}$,$\mathbf{q}$ are the impulsions, $T$ the
temperature and Tr means Trace of the Green function matrix $\hat{G}$ deduced from Eq. (\ref{eq:greenMat4x4}).
\textcolor{black}{Using the relation (\ref{eq:susgen}) we describe
the spin dynamics in pure RES, pure SC phase and coexisting SC-RES
phases.}

\subsubsection{The bare spin susceptibility in the SC phase }

In the pure $d$-wave SC state, the bare susceptibility
writes \cite{Norman07,Schrieffer64}:  
\begin{widetext}
\begin{align}
\chi_{S,sc}^{0}(\omega,\mathbf{q})=\sum_{\mathbf{k}}\left[\frac{1}{2}\left(1+\frac{\xi_{\mathbf{k}}\xi_{\mathbf{k+q}}+\Delta_{SC,\mathbf{k}}\Delta_{SC,\mathbf{k+q}}}{E_{\mathbf{k}}E_{\mathbf{k+q}}}\right)\frac{n_{F}(E_{\mathbf{k+q}})-n_{F}(E_{\mathbf{k}})}{\omega+i\eta-(E_{\mathbf{k+q}}-E_{\mathbf{k}})}\right.\nonumber \\
+\frac{1}{4}\left(1-\frac{\xi_{\mathbf{k}}\xi_{\mathbf{k+q}}+\Delta_{SC,\mathbf{k}}\Delta_{SC,\mathbf{k+q}}}{E_{\mathbf{k}}E_{\mathbf{k+q}}}\right)\frac{1-n_{F}(E_{\mathbf{k+q}})-n_{F}(E_{\mathbf{k}})}{\omega+i\eta+(E_{\mathbf{k+q}}+E_{\mathbf{k}})}\nonumber \\
+\left.\frac{1}{4}\left(1-\frac{\xi_{\mathbf{k}}\xi_{\mathbf{k+q}}+\Delta_{SC,\mathbf{k}}\Delta_{SC,\mathbf{k+q}}}{E_{\mathbf{k}}E_{\mathbf{k+q}}}\right)\frac{n_{F}(E_{\mathbf{k+q}})+n_{F}(E_{\mathbf{k}}-1)}{\omega+i\eta-(E_{\mathbf{k+q}}+E_{\mathbf{k}})}\right],\label{eq:sussc}
\end{align}

\end{widetext}

where $E_{\mathbf{k}}=\sqrt{\xi_{\mathbf{k}}^{2}+\Delta_{SC,\mathbf{k}}^{2}}$ describes the
SC excitations spectrum and $n_{F}$ is the Fermi-Dirac statistic.
\textcolor{black}{The $d$-wave form factor of the SC-order parameter
implies that the coherence factor is maximal on the Fermi surface.
The imaginary part of the bubble exhibits a discontinuity at certain
threshold, and coincidently the real part shows} a logarithmic divergence\textcolor{black}{.
This observation alone enables us to explain in a self consistent
way the formation of the triplet collective mode. Indeed, below this
energy threshold, the divergence in the real part of the spin polarization
cannot be screened by the imaginary part (which vanishes below the
threshold), hence leading to the emergence of the collective mode.
The value of the threshold is expected to be $2|\Delta_{SC}(\mathbf{k}_{HS})|$-
the factor 2 coming from the Green's functions in the bubble, where
$\mathbf{k}_{HS}$ is the momentum of the hotspots. The latter divergence
guarantees the emergence of a collective mode below threshold.} In
order to explain the emergence of a collective mode at ${\bf {Q}}$,
the coherent factors have to be non zero at the Fermi surface while
the FS is gapped. This condition can be fulfilled if we consider a
$d$-wave SC state \cite{Norman07}. This description well reproduces
the imaginary part of the dynamic spin susceptibility inside the SC
state in the overdoped case \cite{Norman07,Eschrig_rev}, with in
particular, the ``X-shape'' form of the dispersion of the modes
around $\left(\pi,\pi\right)$ correctly given in shape and energy
within this simple model. Further we consider that this model gives
a good description of the phenomenon and focus on its generalization
to the PG state.

\subsubsection{The bare spin susceptibility in the RES}

\textcolor{black}{In our theory for the PG state, we evaluate the
spin susceptibility in the RES. The bare spin susceptibility in the RES
writes \cite{Schrieffer64}: } 
\begin{widetext}
\textcolor{black}{{} 
\begin{align}
\chi_{S,RES}^{0}(\omega,\mathbf{q})=\sum_{\mathbf{k}}\left[\frac{1}{4}\left(1+\frac{(\xi_{\mathbf{k}}-\xi_{\mathbf{k}+\pff})(\xi_{\mathbf{k+q}}-\xi_{\mathbf{k+q}+\pffq})+4\Delta_{RES,\mathbf{k}}\Delta_{RES,\mathbf{k+q}}f(\mathbf{q})}{(W_{+,\mathbf{k}}-W_{-,\mathbf{k}})(W_{+,\mathbf{k+q}}-W_{-,\mathbf{k+q}})}\right)\right.\nonumber \\
\left(\frac{n_{F}(W_{-,\mathbf{k}})-n_{F}(W_{-,\mathbf{k+q}})}{\omega+i\eta+W_{-,\mathbf{k}}-W_{-,\mathbf{k+q}}}+\frac{n_{F}(W_{+,\mathbf{k}})-n_{F}(W_{+,\mathbf{k+q}})}{\omega+i\eta+W_{+,\mathbf{k}}-W_{+,\mathbf{k+q}}}\right)\nonumber \\
+\frac{1}{4}\left(1-\frac{(\xi_{\mathbf{k}}-\xi_{\mathbf{k}+\pff})(\xi_{\mathbf{k+q}}-\xi_{\mathbf{k+q}+\pffq})+4\Delta_{RES,\mathbf{k}}\Delta_{RES,\mathbf{k+q}}f(\mathbf{q})}{(W_{+,\mathbf{k}}-W_{-,\mathbf{k}})(W_{+,\mathbf{k+q}}-W_{-,\mathbf{k+q}})}\right)\nonumber \\
\left.\left(\frac{n_{F}(W_{-,\mathbf{k}})-n_{F}(W_{+,\mathbf{k+q}})}{\omega+i\eta+W_{-,\mathbf{k}}-W_{+,\mathbf{k+q}}}+\frac{n_{F}(W_{+,\mathbf{k}})-n_{F}(W_{-,\mathbf{k+q}})}{\omega+i\eta+W_{+,\mathbf{k}}-W_{-,\mathbf{k+q}}}\right)\right],\label{eq:susres}
\end{align}
where $W_{\pm,\mathbf{k}}=\frac{1}{2}\left(\xi_{\mathbf{k}}+\xi_{\mathbf{k}+\pff}\pm\sqrt{(\xi_{\mathbf{k}}-\xi_{\mathbf{k}+\pff})^{2}+4\Delta_{RES,\mathbf{k}}^{2}}\right)$
is the RES excitations spectrum and $f(\bq)$ a function of momentum $\bq$ that takes into account the coherence conditions of the RES , as detailled further in the text and in Appendix \ref{feynman}.} 
\end{widetext}

The contribution to the bare spin susceptibility
in the RES (equation (\ref{eq:susres})) can be divided in two parts
: the intraband contribution (upper terms in relation (\ref{eq:susres}))
and the interband contribution (lower terms in the relation (\ref{eq:susres})).
Close to $\bq=\bbq$, the intraband contribution can be neglected
and the whole signal is produced by interband processes. As the FS
formed by the hybridized bands cannot be connected by the vector $\bbq$,
the bare spin susceptibility is gapped up to the energy $2|\Delta_{RES}(\mathbf{k}_{HS})|$.
Far from $\bq=\bbq$, the intraband processes become non negligible.

Deeper investigation on the SU(2) symmetry have shown that the SU(2) pairing fluctuations emerging from non-linear $\sigma$ model only exist in a restricted area $S_{\kv}$ in the AN part of the first BZ (see Ref. \cite{Montiel_1611} for the detailed demonstration and particularly the figure 9 where $S_{\kv}$ is represented). In the following, one important element is that we assume a symmetrization of this restricted area between two adjacent AN area (in $\kv$  and $\kv+\bbq$ with $\bbq=\left(\pi,\pi\right)$) such that $S_{\kv}=S_{\kv+\bbq}$.

\begin{figure}[h]
\includegraphics[width=7cm]{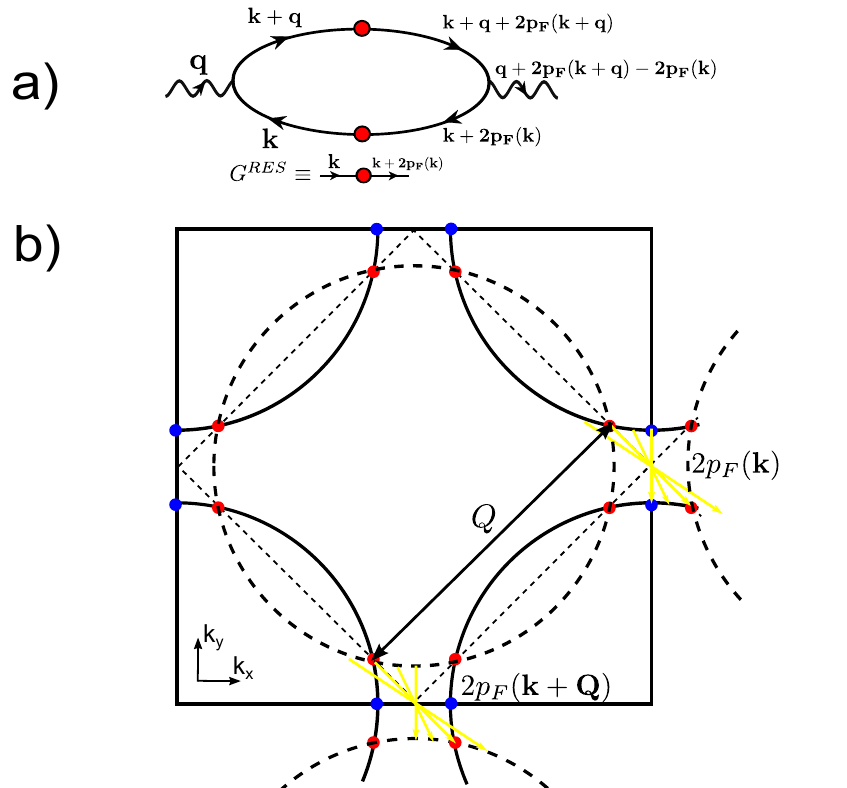} \caption{\label{Figdiag}(Color online) \textcolor{black}{a) Diagrammatic contribution
that describes the coherence between two particle-hole patches. The
outgoing vector depends on the difference $2p_{F}(\mathbf{k+q})-2p_{F}(\mathbf{k})$
which vanishes at commensurate $\bq$ vectors. This Feynman diagram
must vanish for incommensurate $\bq$ vectors and exists close to
commensurate $\bq$ vectors. b) Representation of the Fermi surface
of the electrons at ${\bf {k}}$ (solid line) and the electrons at
${\bf {k+Q}}$ (dashed line). The $2p_{F}$ vectors of the electrons
at ${\bf {k+Q}}$ are the same as the electrons at ${\bf {k}}$
(yellow arrows).}}
\end{figure}

The coherence terms are described by the Feynman
diagram shown in Fig. \ref{Figdiag} a)
and we observe that the outgoing vector of the Feynman diagram does not equal the incoming
vector $\mathbf{q}$ up to the difference $\bar{\delta}_{\pf}=\pffq-\pff$.
The difference $\bar{\delta}_{\pf}$ vanishes ($\bar{\delta}_{\pf}=\mathbf{0}$) only if $\mathbf{q}$ is commensurate and differs from zero
($\bar{\delta}_{\pf}\neq \mathbf{0}$) for incommensurate $\mathbf{q}$ vectors (see Fig. \ref{Figdiag}).
Consequently the coherence terms exist only close $\bf{q}=0,\bf{Q}$ and cannot exist far from commensurate vectors. In the following, we modelize the RES coherence terms in Eq. (\ref{eq:susres}) by the the terms $\Delta_{RES,\mathbf{k}}\Delta_{RES,\mathbf{k}+\bq}f(\bq)$ where $f(\bq)$ vanishes for incommensurate $\bq$ vectors. More precisely, the function $f(\bq)$ equals one around $\mathbf{q}=\mathbf{0}$ and $\mathbf{q}=\mathbf{Q}$ and vanishes for other vectors. A full description of the function $f(\bq)$ is done in the appendix \ref{feynman} while the effect of the function $f(\bq)$ on the spin susceptibility is studied in appendix \ref{feffect}.

In contrast to the preformed Cooper pair scenario \cite{Norman07}
we observe a resurgence of the coherence terms around incoming wave
vectors commensurate with the lattice, like $\mathbf{q}=\mathbf{Q}$.
In the RES scenario, the coherence terms only exist close to commensurate $\bq$ vectors.
This peculiar behavior is different from the scenario of preformed
Cooper pairs where the coherence terms vanish for all $\bq$ vectors.

Close to the FS, we can linearize the shifted spectrum $\xi_{\mathbf{k}-\pff}$. This linearization
leads to the relation $\xi_{\mathbf{k}-\pff}\approx-\xi_{\mathbf{k}}$
only valid close to the FS. In this approximation, the relation (\ref{eq:susres})
is equal to the relation (\ref{eq:sussc}). We can deduce that the
low energy spectrum in the RES and the SC state are nearly the same.

\subsubsection{The bare spin susceptibility in the coexisting
SC+RES phase}

\textcolor{black}{A detailed study of the whole Feynman diagram that
contributes to the bare spin susceptibility is done in Appendix \ref{feynman}.
The main contribution of the SC and the RES state does not qualitatively
change regarding the pure state study. The threshold in the bare spin
susceptibility occurs at an energy $2\sqrt{\Delta_{RES}^{2}(\mathbf{k}_{HS})+\Delta_{SC}^{2}(\mathbf{k}_{HS})}$
and depends on both SC and RES state. In addition to the RES coherent
terms, note that a mixed SC+RES exists and also contributes only close
to commensurate ${\bf {q}}$ vector (see Appendix \ref{feynman})}

\subsection{Raman response function}

\textcolor{black}{The Raman Response $\chi_{\lambda}$ is the response
function of a modified density operator $\chi_{\lambda}=-i\Theta(\tau)\langle\rho^{\lambda}(\tau)\rho^{\lambda}(0)\rangle$
with $\rho^{\lambda}=\sum_{{\bf {k}}}\gamma^{\lambda}c_{{\bf {k}}}^{\dagger}c_{{\bf {k}}}$
where $\gamma^{\lambda}$ is the Raman vertex in the symmetry $\lambda$
\cite{cardona97,Devereaux-RMP}. The Raman susceptibility strongly
depends on the symmetry of the system. We can take into account these
symmetries by considering vertices in the phonon-matter interaction
different from unity. In cuprates compounds, we typically study three
symmetries which are written within the effective mass approximation:
\begin{align}
\gamma^{B_{1g}}=\frac{1}{2}\left[\frac{\partial^{2}\xi_{k}}{\partial k_{x}^{2}}-\frac{\partial^{2}\xi_{k}}{\partial k_{y}^{2}}\right]\nonumber \\
\gamma^{B_{2g}}=\frac{1}{2}\left[\frac{\partial^{2}\xi_{k}}{\partial k_{x}\partial k_{y}}+\frac{\partial^{2}\xi_{k}}{\partial k_{y}\partial k_{x}}\right]\nonumber \\
\gamma^{A_{1g}}=\frac{1}{2}\left[\frac{\partial^{2}\xi_{k}}{\partial k_{x}^{2}}+\frac{\partial^{2}\xi_{k}}{\partial k_{y}^{2}}\right]
\end{align}
were the $B_{1g}$ symmetry that probes the AN zone of the first BZ,
the $B_{2g}$ symmetry probes the N zone of the first BZ And the $A_{1g}$
symmetry probes the whole Brillouin zone. Here, we do not consider
the $A_{2g}$ symmetry, $\gamma^{A_{2g}}=0$. In the following, we
only focuses on the $B_{1g}$ and $B_{2g}$ symmetry. The specific
case of $A_{1g}$ symmetry has already been studied in the framework
of a charge order and superconducting coexisting state \cite{Montiel15a}.
In both the $B_{1g}$ and the $B_{2g}$ symmetries, the Coulombian
screening can be neglected \cite{cardona97}. In the $B_{1g}$ and
$B_{2g}$ symmetries, the bare Raman susceptibility write \cite{cardona97,Devereaux-RMP}:
\begin{align}
\chi_{\lambda}\left(\omega,\mathbf{q}=0\right)=-\frac{T}{2}\sum_{\epsilon,\mathbf{k}}\text{Tr}\left[\bar{\gamma}^{\lambda}(\mathbf{k})\hat{G}\left(\omega+\varepsilon,\mathbf{k}\right)\bar{\gamma}^{\lambda}(\mathbf{k})\hat{G}\left(\varepsilon,\mathbf{k}\right)\right],\label{eq:susram}
\end{align}
where $\bar{\gamma}^{\lambda}(\mathbf{k}=\gamma^{\lambda}(\mathbf{k})\bar{\tau_{3}}$
with $\tau_{3}$ is the Pauli matrix evolving in the particle-hole
space in the $\lambda$ symmetry (with $\lambda=B_{1g}$ or $B_{2g}$). }

\section{Results\label{results} }

We perform a study at optimal doping $p=0.16$ and
in the underdoped regime $p=0.1$ in Hg-1201. At both $p=0.1$ and
$p=0.16$, the SC critical temperature $T_{c}$ is lower than the
$T^{*}$, $T_{c}<T^{*}$. We consider that in the SC state $T<T_{c}$,
the SC and RES coexist while above $T_{c}$ ($T_{c}<T<T^{*}$) only
the RES remains. The RES disappears at $T^{*}$.

At $p=0.1$, we choose $\Delta_{RES}^{0}=0.09eV$
and $\Delta_{SC}^{0}=0eV$ in the RES state while $\Delta_{RES}^{0}=0.06eV$
and $C^{0}=0.06eV$ in the SC state. The order magnitude of the RES
and SC order parameter on the Fermi surface is presented in Fig. \ref{densitep10}
a) and b). The SC order parameter develops in the N region and decreases
in the AN zone while the RES order parameter vanishes in the N region
and increases in the AN region. At the zone edges, the SC gap represents
$30\%$ of the whole gap magnitude while the RES is at $70\%$.

At optimal doping ($p=0.16$), we choose $\Delta_{RES}^{0}=0.065eV$
and $\Delta_{SC}^{0}=0eV$ in the RES state while $\Delta_{RES}^{0}=0.01eV$
and $C^{0}=0.042eV$ in the SC state as presented on Fig. \ref{densitep16}.
The SC order parameter develops on the whole Fermi surface while the
RES order parameter only exists in the AN zone. At the zone edges,
the SC gap represents $95\%$ of the whole gap magnitude while the
RES is at $5\%$. The SC order parameter exhibits a $d$-wave aspect
at optimal doping while this aspect is weaken in the underdoped regime.
The RES gap dependence is different than a pure $d$-wave dependence
as observed by ARPES in Bi-2212 \cite{Vishik12} and Hg-1201 \cite{Vishik14_He}.

From a technical point of view, the calculation of the bare polarization
bubbles is done as follow. The summation over the internal impulsion
is done in a 400x400 grid in the first BZ after doing the analytical
integration over the internal Matsubara frequencies at $T=0K$. Note
that we neglected the temperature dependence of the order parameters.
We have done the analytical continuation on the external Matsubara
frequency replacing $i\omega$ by $\omega+i\eta$ where $\eta$ is
a small damping parameter taken here to $\eta=3\,meV$. This small
parameter can be understood as residual scattering caused by the impurities.
The susceptibilities are in the unit of states per eV per CuO$_{2}$
formula unit and should be multiplied $2\mu_{B}^{2}$ to compare to
neutron-scattering data ($\mu_{B}$ is the Bohr magneton).

\subsection{\textcolor{black}{The density of states.}}

\begin{figure}[h]
\includegraphics[width=8cm]{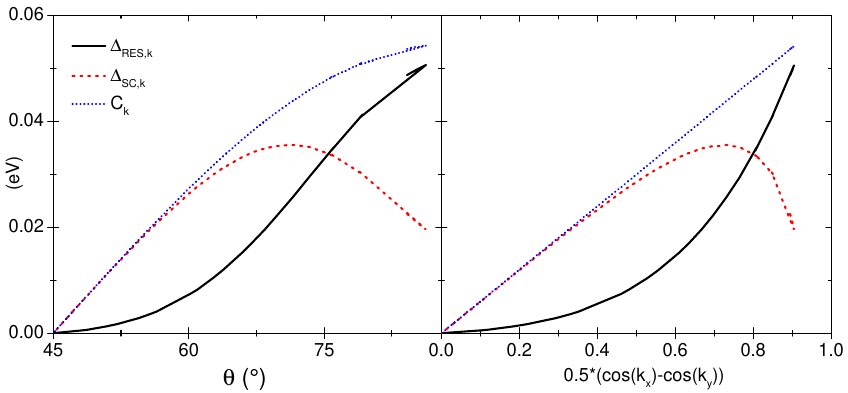} 
\includegraphics[width=8cm]{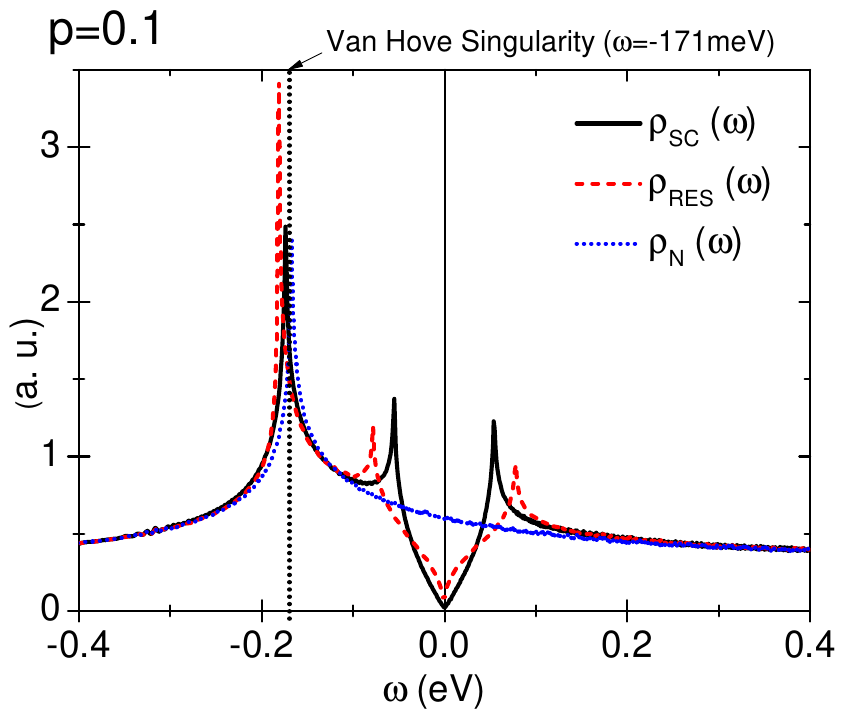}
\caption{\label{densitep10} Dependence of the RES (solid line), SU(2) envelop
(dotted line) and SC gap (dashed line) on the FS as a function of
left panel) the $\theta$ angle and right panel) the $d$-wave factor
at $p=0.1$. The SC gap exhibits a $d$-wave behavior close to the
nodal zone and its intensity decreases in the AN zone. Bottom panel)
Density of states in the Normal metal $\rho^{N}(\omega)$ (dotted
lines), the RES $\rho^{RES}(\omega)$ (solid lines) and the SC state
$\rho^{SC}(\omega)$ (dashed line) as a function of energy $\omega$
for $p=0.1$. The SC and RES order parameters open a symmetric gap
centered around the Fermi level $\omega=0\,eV$. The $d$-wave symmetry leads to the typical form of the density of state at low energy. The Van-Hove singularity
arises in the metallic spectrum at $\omega=-171meV$ for $p=0.1$.
The magnitude of the gap is $54\,meV$ in the SC phase and $75\,meV$
in the RES phase. Note that the amplitude of the gaps are qualitatively
in accordance with the experimental data.}
\end{figure}

\begin{figure}[h]
\includegraphics[width=8cm]{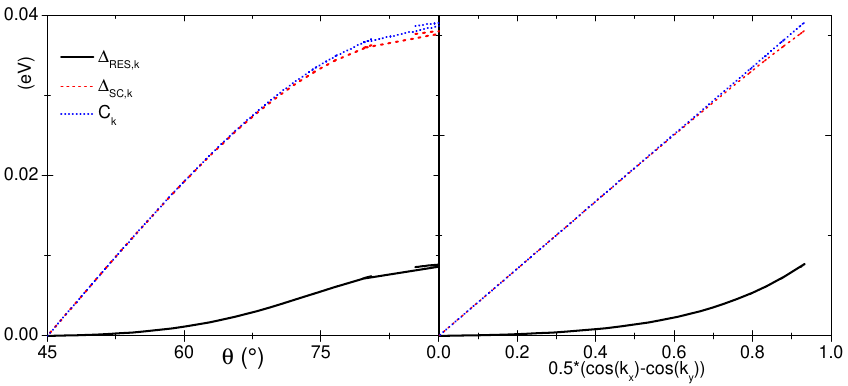}
\includegraphics[width=8cm]{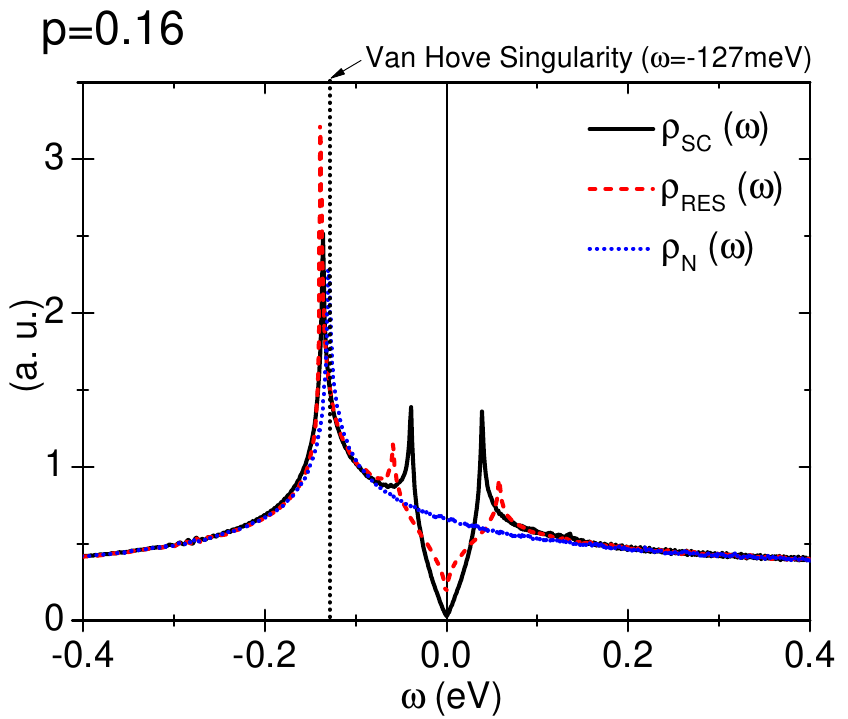}
\caption{\label{densitep16} Dependence of the RES (solid line), SU(2) envelop
(dotted line) and SC gap (dashed line) on the FS as a function of
left panel) the $\theta$ angle and right panel) the $d$-wave factor
at $p=0.1$. The SC order parameter as $d$-wave behavior in the whole
Brillouin zone. Bottom panel) Density of states in the Normal metal
$\rho^{N}(\omega)$ (dotted lines), the RES $\rho^{RES}(\omega)$
(solid lines) and the SC state $\rho^{SC}(\omega)$ (dashed line)
as a function of energy $\omega$ for $p=0.16$. The SC and RES order
parameters open a symmetric gap centered around the Fermi level $\omega=0\,eV$.
The $d$-wave symmetry leads to the typical form of the density of state at low energy.
The Van-Hove singularity arises in the metallic spectrum at $\omega=-292meV$
for $p=0.16$. The magnitude of the gap is $39\,meV$ in the SC phase
and $59\,meV$ in the RES phase. Note that the amplitudes of the gaps
are qualitatively in accordance with the experimental data.}
\end{figure}

The electronic density of states (DOS), $\rho(\omega)=\frac{-2}{\pi}\sum_{\mathbf{k}}\left[\text{Im}\left(\text{lim}_{\eta\rightarrow0}G^{11}(\omega+i\eta,\mathbf{k})\right)\right]$
in the normal metal, RES and SC phases are plotted in Fig. \ref{densitep10}
for hole doping $p=0.1$ and Fig. \ref{densitep16} for $p=0.16$.
Both SC and RES open a symmetric gap at the Fermi level ($\omega=0$).
At $p=0.1$, the magnitude of the gap is $54\,meV$ in the SC phase
and $75\,meV$ in the RES phase. At $p=0.16$, the magnitude of the
gap is $39\,meV$ in the SC phase and $59\,meV$ in the RES phase.
The amplitude of the gap in the RES and the SC state are in good agreement
with experimental gaps deduced from Raman scattering\cite{LeTacon108,LeTacon111}.
The low energy behavior of the DOS differs a little between the RES
and the SC state. The coherent peak seen in the SC state is weakened
in the RES state as observed in cuprate compounds \cite{Fisher07}.
Note that the Van Hove singularity is well defined by a peak at negative
energy.
The from of the DOS at low energy (close to $\omega=0\,eV$) is typical of the $d$-wave momentum dependence of the SC gap \cite{Fisher07} but does not give more information about the nature of the order parameter. In order to observe specific signature of both RES and SC state, we need probes that are sensible to the coherence between
the quasiparticles such as Raman scattering and INS.

\subsection{The Raman susceptibility}

\begin{figure}[h]
\includegraphics[scale=0.48]{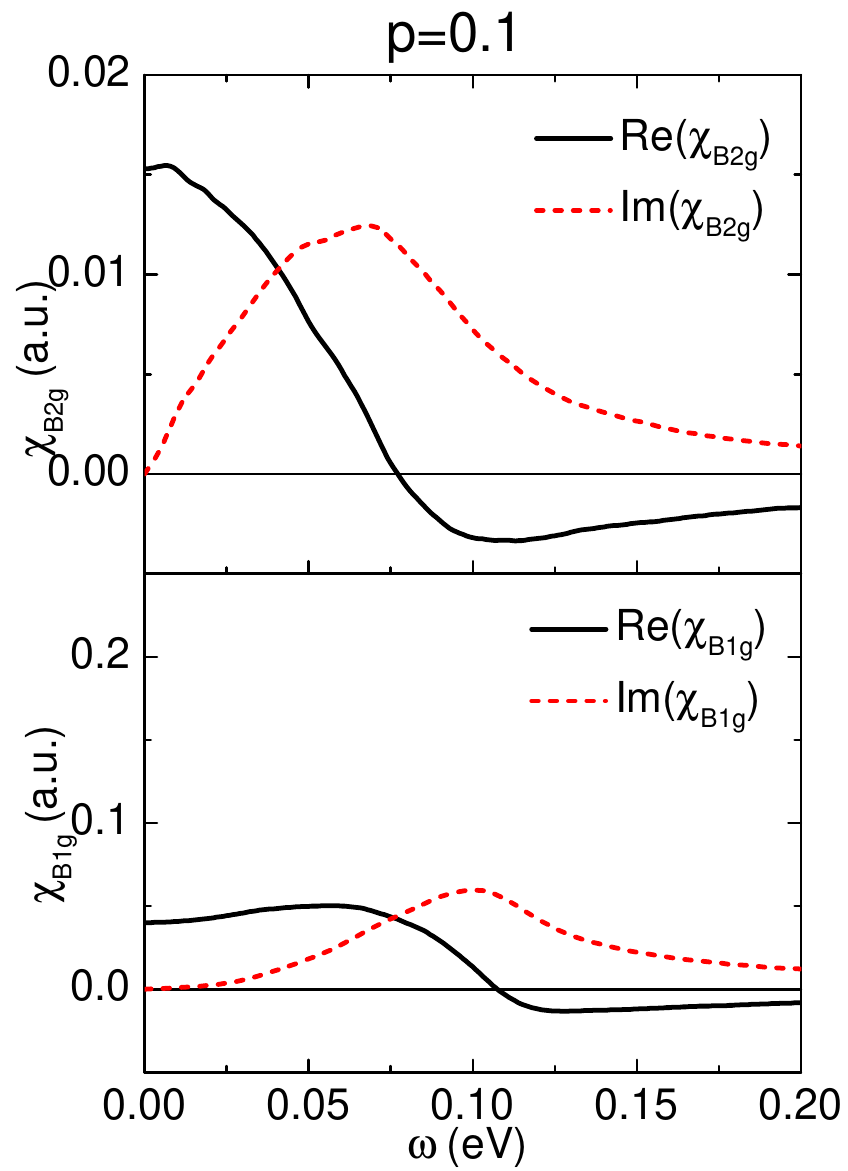} \includegraphics[scale=0.48]{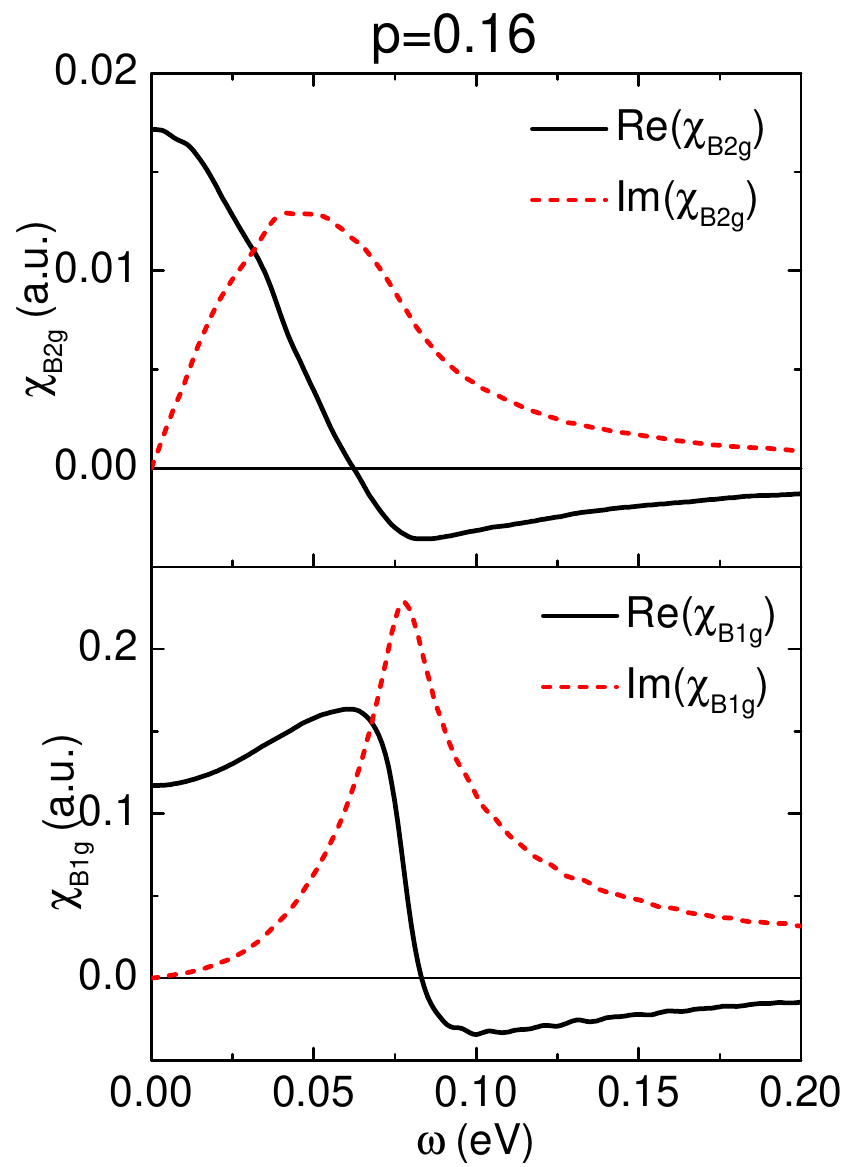}
\caption{\label{bareRaman} The Raman susceptibility in the $B_{1g}$ and the
$B_{2g}$ symmetry at a) p=0.1 in the RES state, b) and b) p=0.16.
The calculated responses are in promising agreement with the experimental
data. In the $B_{2g}$ channel, the intensity does not vary with doping
but the frequency resonance increases at low doping. In the $B_{1g}$
channel, the frequency resonance increases at low doping but the intensity
fall down as observed experimentally.}
\end{figure}

We calculate the Raman response in the $B_{1g}$
and the $B_{2g}$ symmetry in the SC state at $p=0.1$ and $p=0.16$
(see Fig. \ref{bareRaman}) \cite{cardona97,Devereaux-RMP}. Our approximation
is able to reproduce the decreasing of the frequency resonance in
the $B_{1g}$ symmetry with hole doping, (see Fig. \ref{bareRaman})
from $\omega_{sc}=101\,meV$ at $p=0.1$ until $\omega_{sc}=77\,meV$
at $p=0.16$. Moreover, the intensity of the $B_{1g}$ Raman resonance
is lower at low doping (p=0.1) than close to optimal doping (p=0.16).
Both features are in good agreement with experimental Raman scattering
in Hg-1201 compound \cite{LeTacon108,LeTacon111}.

In the $B_{1g}$, the superconducting coherence
peak occurs at the energy $2\sqrt{\Delta_{RES}^{2}(\mathbf{k}_{ZE})+\Delta_{SC}^{2}(\mathbf{k}_{ZE})}$
where $\mathbf{k}_{ZE}$ is the point of the FS localized at the zone
edge (see Fig. \ref{FS}). The frequency of the superconducting coherent
peak depends on the magnitude of both the SC and RES order parameters
at the zone edge. Consequently, this frequency is larger than twice
the magnitude of the SC order parameter and does not scale with $T_{c}$.
However, the intensity of the SC coherent peak only depends on the
magnitude of the SC order parameter at the zone edge. In step with
the SC gap dependence discussed in section \ref{min} and shown in
Figs. \ref{densitep10} and \ref{densitep16}, the intensity of the
SC coherent peak in the $B_{1g}$ symmetry increases with the hole
doping.

In the $B_{2g}$ channel, we see the emergence of
a peak at low frequency \cite{LeTacon108,LeTacon111}. The $d$-wave
symmetry of the gap implies a small intensity of the SC coherent peak.

\subsection{The bare spin susceptibility}
\begin{verse}
\begin{figure}[h]
\includegraphics[scale=0.48]{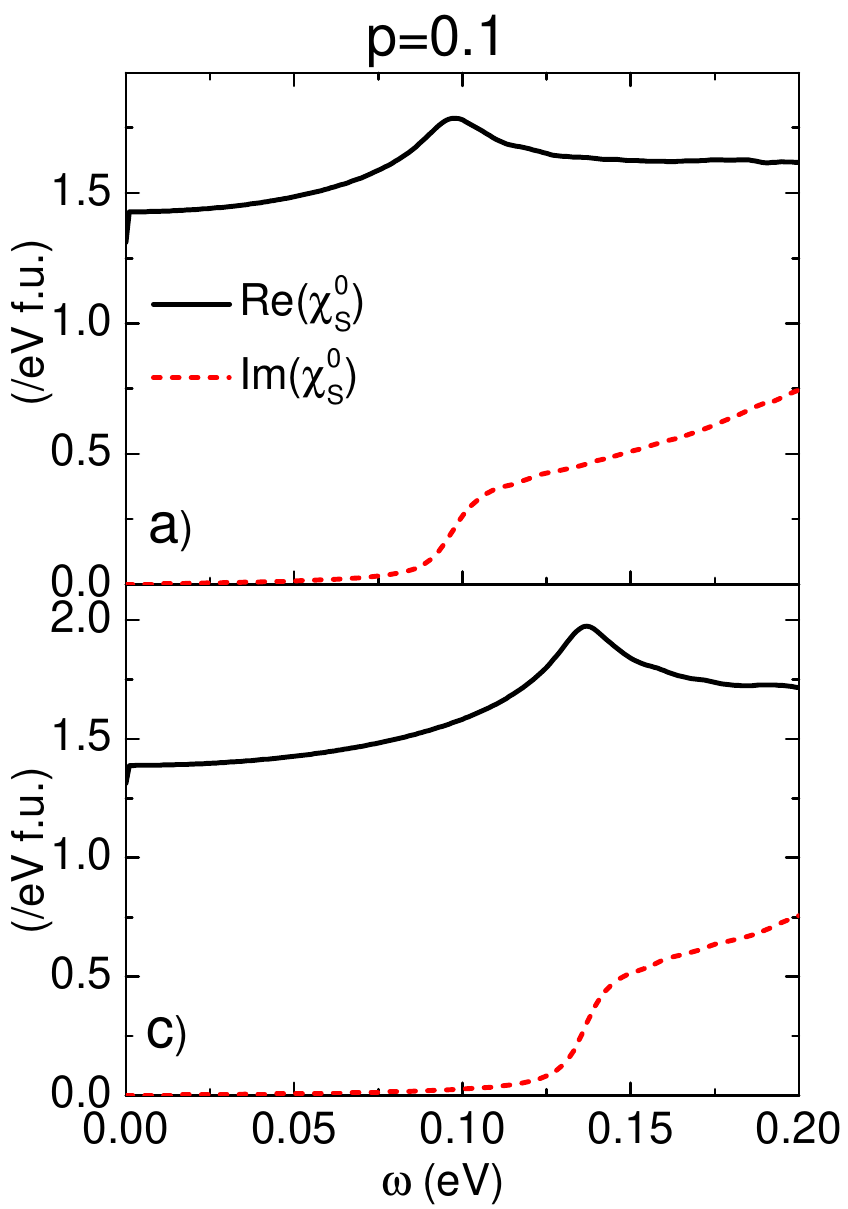} \includegraphics[scale=0.48]{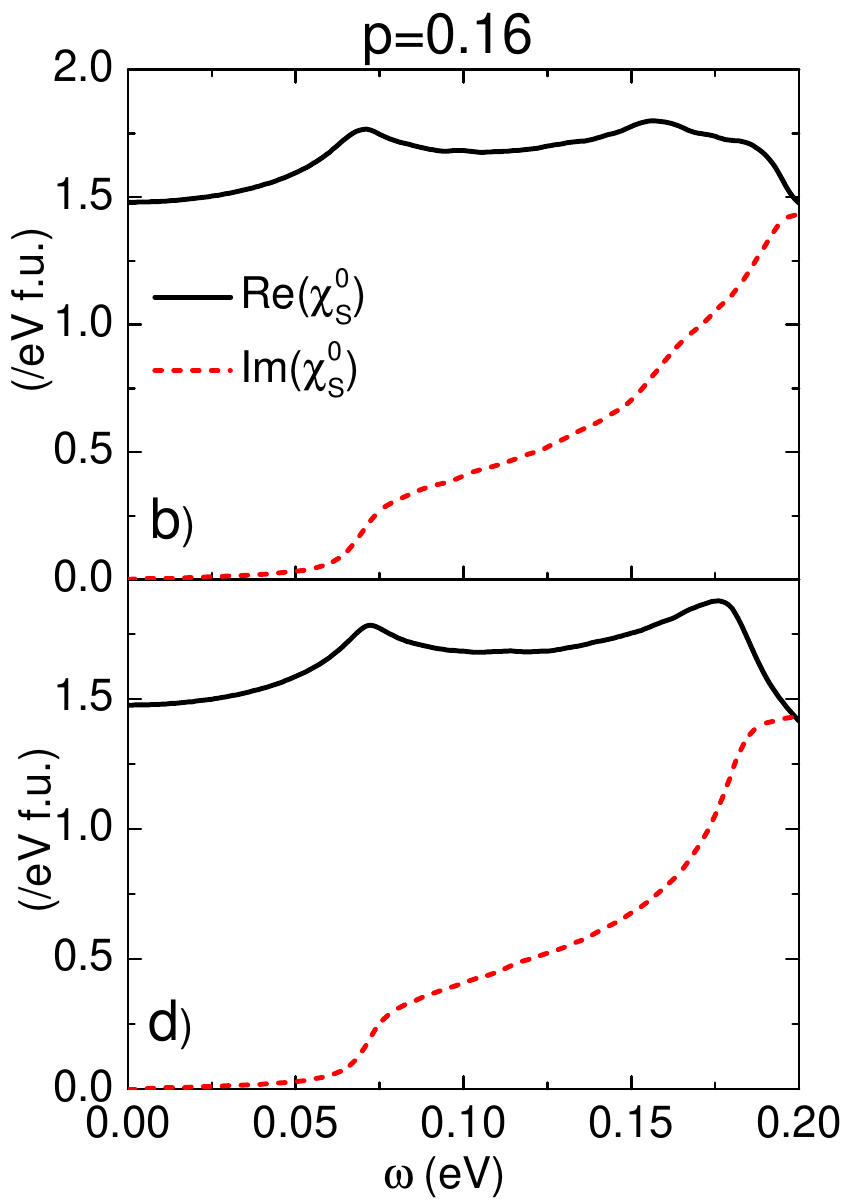}
\caption{\label{bareRES} The real and imaginary parts of the bare polarization
bubble at ${\bf {Q}=(\pi,\pi)}$ for a) $p=0.1$ in the RES state,
b) $p=0.16$ in the RES state c) $p=0.1$ in the SC state, d) $p=0.16$
in the SC state. The amplitude of the order parameters are the same
as in Figs. \ref{densitep10} and \ref{densitep16}. We observe
a gap opening in both SC state and RES. }
\end{figure}

\end{verse}
The real and imaginary parts of the bare polarization
bubble in the RES and SC phases at hole doping $p=0.1$ and $p=0.16$
are presented in Fig.\ref{bareRES} as a function of $\omega$ at
${\bf {Q}=(\pi,\pi)}$. In the RES, (Figs. \ref{bareRES} a) and b)),
a gap opens in the imaginary part of $\chi_{S}^{0}$
very similarly than the quasiparticle gap opening in the SC state
(Figs. \ref{bareRES} c) and d)). In the RES, the threshold in the
Imaginary part and the logarithmic divergence in the real part occur
at energies close to $2\Delta_{RES}(\mathbf{k}_{HS})$. The energy
of the threshold move from $94\,meV$ at $p=0.1$ until $64\,meV$
at $p=0.16$ in the RES. On the other hand, the threshold is defined
at the energy $2\sqrt{\Delta_{RES}^{2}(\mathbf{k}_{HS})+\Delta_{SC}^{2}(\mathbf{k}_{HS})}$
in the SC phase. The energy of the threshold moves from $129\,meV$
at $p=0.1$ down to $66\,meV$ at $p=0.16$ in the SC. The bare spin
susceptibilities in the SC and RES states are very similar because
the gap mechanism is nearly the same close to the FS. This feature
is emphasized by the fact that close to the FS, we can apply the identity
$\xi_{{\bf {k}-\pff}}=-\xi_{{\bf {k}}}$ and the bare spin susceptibility
in the RES becomes the same than the SC one.

\subsection{The RPA spin susceptibility}

\begin{figure}[ph]
a)\includegraphics[width=8cm]{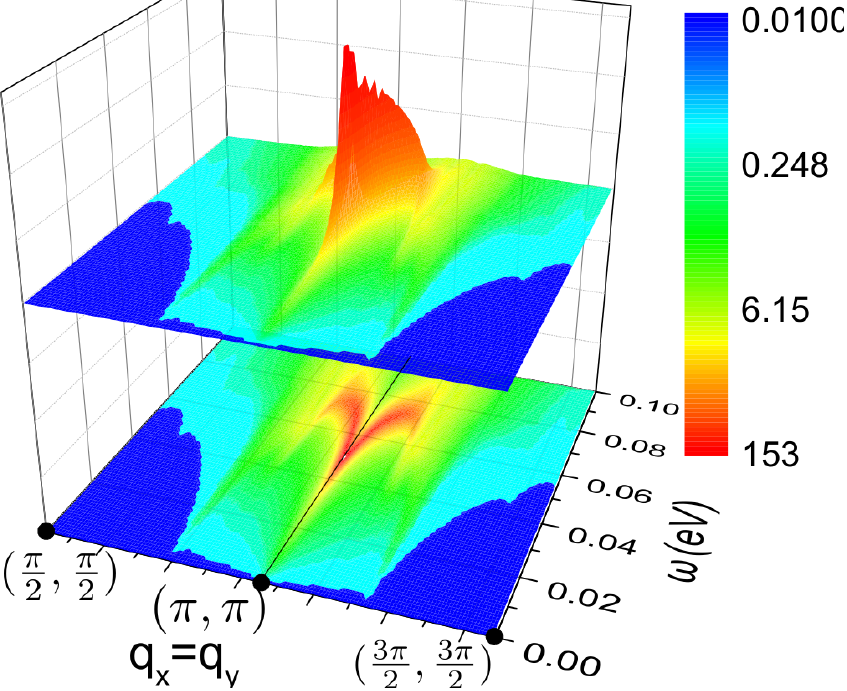} 

b)\includegraphics[width=8cm]{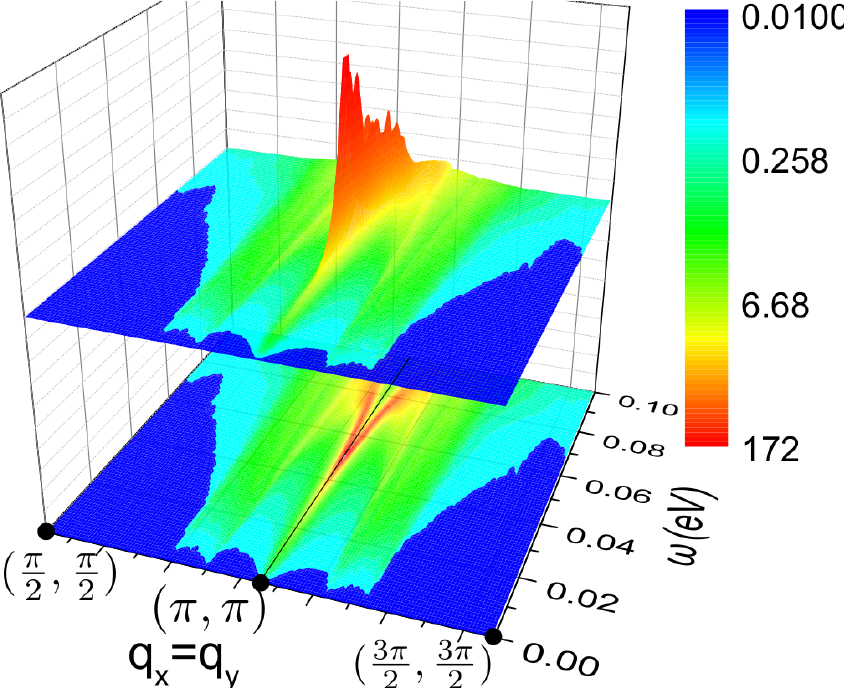} 

c)\includegraphics[width=8cm]{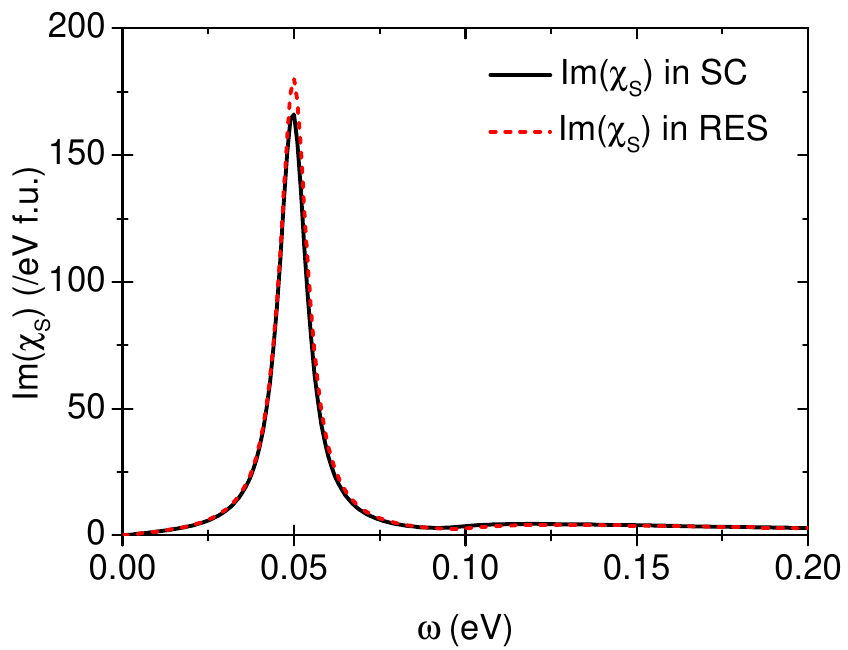} \caption{\label{RPARESp10} Amplitude of the Imaginary part of the spin susceptibility
$\chi_{S}$ as a function of $\omega$, for $q_{y}=q_{x}$ and $q_{x}$
from $-\pi/2a$ to $3\pi/2a$ at $p=0.1$ for $J_{0}=169\,meV$ and
$V=100$ in a) the SC state and b) the RES. The solid line is set
at $q=(\pi/a,\pi/a)$. c) Cut at ${\bf {Q}=(\pi,\pi)}$ of the imaginary
part of $\chi_{S}$ in the RES (dashed line) and SC (solid line) state.}
\end{figure}

\begin{figure}[ph]
a)\includegraphics[width=8cm]{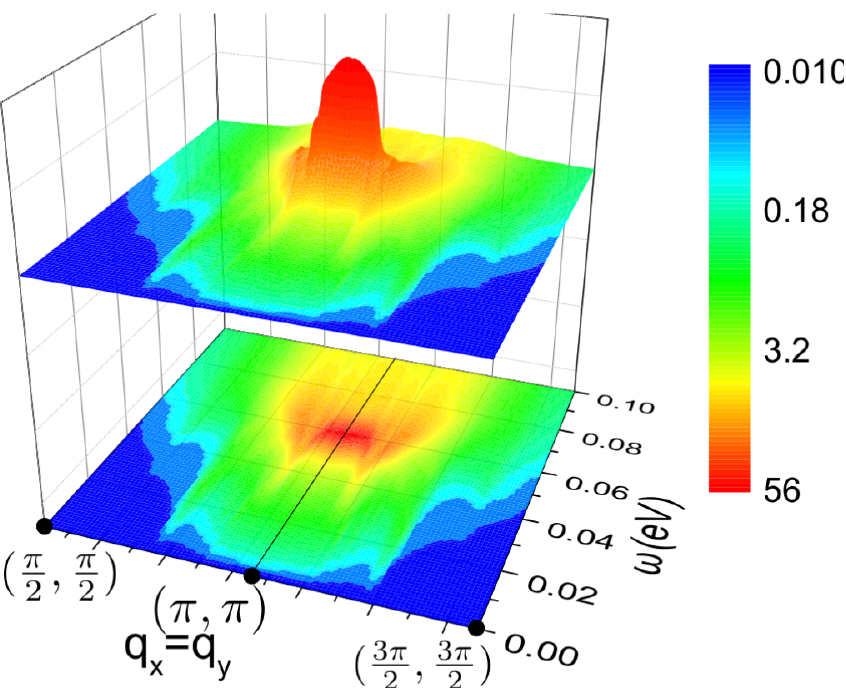} 

b)\includegraphics[width=8cm]{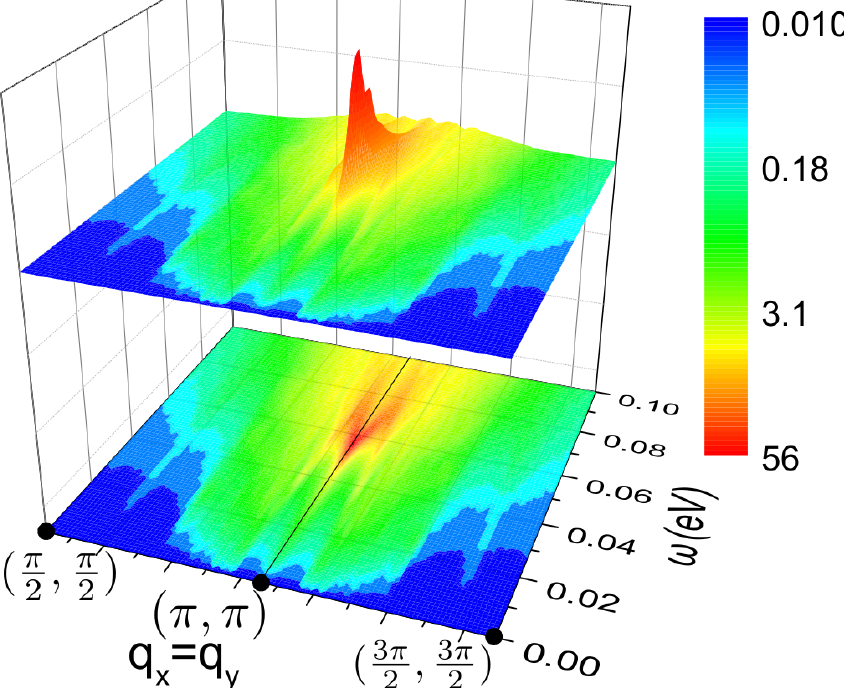} 

c)\includegraphics[width=8cm]{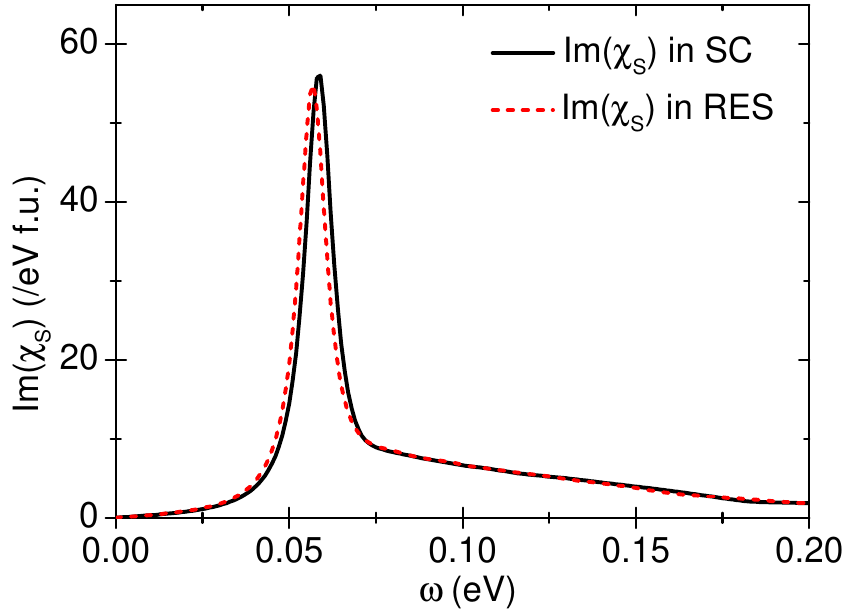} \caption{\label{RPARESp16} Amplitude of the Imaginary part of the spin susceptibility
$\chi_{S}$ as a function of $\omega$, for $q_{y}=q_{x}$ and $q_{x}$
from $-\pi/2a$ to $3\pi/2a$ at $p=0.16$ for $J_{0}=151\,meV$ and
$V=100$ in a) the SC state and b) the RES. The solid line is set
at $q=(\pi/a,\pi/a)$. c) Cut at ${\bf {Q}=(\pi,\pi)}$ of the imaginary
part of $\chi_{S}$ in the RES (dashed line) and SC (solid line) state.}
\end{figure}

The amplitude of the imaginary part of the RPA susceptibility for
the RES and SC phases are presented in Figs. \ref{RPARESp10} and
\ref{RPARESp16} for $p=0.1$ and $p=0.16$) respectively, as a function
of $\omega$ in the diagonal direction $q_{y}=q_{x}$ and $q_{x}$
from $-\pi/2a$ to $3\pi/2a$ (with $a$ that the unit cell parameter
set to unity).

At $p=0.16$, the magnitude of the super-exchange interaction $J_{0}=151\,meV$
is adjusted to set the resonance at $\bbq$ at $60\,meV$ while at
$p=0.1$, we put $J_{0}=169\,meV$ to ensure a resonance at $50\,meV$.
\textcolor{black}{In both RES and SC state, we observe a resonance
at $\bbq$. The intensity as well as the form of the resonance does
not vary a lot between the two states at both doping.(Figs. \ref{RPARESp10}
c) and \ref{RPARESp16} c).} The shape of the energy fluctuations
close to $\bbq$ does not qualitatively change between the SC and
RES at $p=0.1$ (Figs. \ref{RPARESp10} a) and b)) while this change
is strong at optimal doping (Figs. \ref{RPARESp16} a) and b)). The
change in the form is a clear effect of the loss of coherence between
the patches away from the $\bbq$ vector in the RES.

At $p=0.1$, the RES order parameter is dominant in both SC and RES
state resulting on a Y-shape in both cases. At optimal doping, the
SC order parameter dominates in the SC states leading to the X-shape.
The loss of coherent terms in the RES erases the X-shape observed
in the SC state.

\section{Discussion\label{discussion} }

We discuss below the three main findings of our theory, compared to
other approaches proposed so far.

First, as shown in Figs. \ref{RPARESp10} and \ref{RPARESp16} the
model gives a good agreement for the frequency resonance and the intensity
of the resonance observed at $\bbq$ in Hg-1201. The frequency resonance
is determined by the values of the RES and SC order parameters as well
as the value of the super-exchange $J_{0}$. The values of the RES
and SC order parameter have been determined to reproduce the Raman
coherent peak in the $B_{1g}$ and $B_{2g}$ symmetry. The value of
$J_{0}=169\,meV$ at $p=0.1$ and $J_{0}=151\,meV$ at $p=0.16$ is
in the right range of values for cuprate compounds. Moreover, the
decrease of the magnitude with doping is consistent with the decrease of the two-magnon peak in Raman data \cite{LeTacon108,LeTacon111}. The frequency and intensity of
the resonance is the same in both the SC and the RES in the underdoped
regime and optimal doping \cite{Vignolle07,Hinkov07,Tranquada09}.
At optimal doping, the same intensity observed in both RES and SC
state is in agreement with the absence of any signature on the intensity
of the resonance at $T_{c}$ \cite{Vignolle07,Hinkov07,Tranquada09}.
The same intensity in both the RES and SC state is a by-product of
our model where we did not adjust the damping that could be higher
in a non-homogeneous state as RES. Indeed, our model produces naturally intrinsical inhomogeneities due to the proliferation of local objects. This aspect will be studied in future publications.

Second, our model reproduces in a promising agreement the fluctuation
spectrum around $\bbq$ in both the SC and RES state. The disappearance
of the low energy fluctuation spectrum in the SC state when we pass
to the RES (and then the transformation from the X-shape to the Y-shape)
can be explained by the loss of the coherence terms in the RES away
from the vector $\bbq$. The enhancement of the
coherence close to the $\bbq$ vector leads to the increasing of the
value of the spin susceptibility at $\bbq$ in the RES and the emergence
of the Y shape in the energy fluctuation spectrum. In our model, we
have modeled this loss of coherence by a function $f(\bq)$ which vanishes
away from $\bbq$. The effect of the width of the function $f(\bq)$
on the spin susceptibility is studied in the Appendix \ref{feffect}.

A simple explanation for the emergence of spectral weight at $\bbq$
in the pseudo gap phase can be given as follows. Since their origin
lies in the SU(2) fluctuations, the RES patches are acting on a small
part of the BZ, and are gapping out the anti-nodal region of the Fermi
surface, close to the hot spots. Fluctuations associated with the
SU(2) scenario are thus restricted to these regions. The typical wave
vectors connecting these regions to one another are $\mathbf{q}=\mathbf{\bbq}$
and $\mathbf{q}=\mathbf{0}$, but due to the presence of the $d$-wave
phase factor, the positive sign necessary for forming bound state
(as opposed to anti-bound) selects the wave vector $\bbq$. Hence
the two main ingredients for the emergence of spectral weight at $\bbq$
and the presence of the factor $f(\bq)$ in Eqn.(\ref{eq:susres})
are the localization of the RES around the hot spots (which selects
the mode modulation vectors $\bq=0$ and $\bq=\bbq)$ in the anti-nodal
region and retaining a certain coherence with $d$-wave form factor (
which finally selects the modulation vector around $\bq=\bbq$). In
order to test this idea, we show in Appendix \ref {gapeffect} the same calculation for a SC state with the SC gap formation restricted
to a small region around the hot spots. We see in Fig. \ref{DepSCgap} that it gives some additional spectral weight around
$\bq=\bbq$ as desired. For a SC state, the form of the additional
spectral weight is more like a spot rather than the ``Y''-shape.
The elongation of the tail of the ``Y'' at $\bbq$ is a consequence
of the ``nesting'' feature $\mathbf{k}\rightarrow\mathbf{k}-\pf$
when the energy is lowered.

Lastly, the dependence in doping of the fluctuation spectrum can be
explained by the nature of the RES. The proliferation of excitonic
patches occurs at zero temperature at low doping while it occurs at
much higher temperature close to optimal doping (see Fig. \ref{phased}).
This difference implies that RES order is strong at low doping and
coexist with SC order parameter while it weakens at optimal doping
consistently with Raman experiments \cite{LeTacon108,LeTacon111}.
Consequently, the RES state drives the physics close to AF critical
vector at low doping explaining the Y-shape of energy fluctuation
spectrum in both SC and RES. At optimal doping, the RES weakens in
the AN zone and the physics is dominated by SC order parameter which
implies the appearance of the X-shape.

A possible extension of this work should be the
calculation of the RES response in bilayered systems. In such systems,
the interlayer coupling creates bonding and anti-bonding states and
gives rise to even and odd spin susceptibilities. Leaving aside the
stability of the RES in such bilayer compounds, we expect the even
and odd susceptibilities to behave similarly than in the monolayer
compound. However, the exact vector where the resonance occurs could
change because of the mismatch between the bonding and anti-bonding
Fermi surfaces. The effect of exotic structure like CuO chains in YBCO compounds on the spin dynamics is still unclear. The CuO chains could stabilize nematic orders \cite{Orth2017} which could reciprocally affect of spin susceptibility and produce incomensurability. This nematic order could be  strong in the SU(2) scenario \cite{Montiel_1611}. We let the detailed calculations for forthcoming publications.

\section{Conclusion\label{conclusion} }

We proposed a description of the energy spectrum
of the dynamic spin susceptibility, observed by INS in recent experiments
on the cuprate compounds Hg-1201, for both the SC and the PG states.
This explanation is based on a new concept for the PG phase which
shows the emergence of parrticle-hole pairs, forming excitonic droplets, or
patches with multiple modulation wave vectors $\pf.$ The RES state
behaves ``almost'' like a $d$-wave SC, but gaps out the anti-nodal
region of the first BZ, leading to the formation of Fermi ``arcs''\cite{Montiel16ARPES}. In the PG regime, this restriction provokes a loss of coherence terms except at some peculiar wave
vectors commensurate with the lattice, like the AF vector $\bbq$.
This description of the PG phase is able to reproduce the main features
of the Raman scattering in Hg-1201, and is a promising candidate for
PG state of superconducting cuprates.
\begin{acknowledgments}
The authors acknowledge Y. Sidis, P. Bourges, S. Hayden, M. Greven and T. Kloss for fruitful
discussions. This work was supported by LabEx PALM (ANR-10-LABX-0039-
PALM), of the ANR project UNESCOS ANR-14-CE05-0007, the ERC, under
grant agreement AdG-694651-CHAMPAGNE, the Aspen Center for Physics,
as well as the Grant No. Ph743-12 of the COFECUB which enabled frequent
visits to the IIP, Natal. X.M. also acknowledge the support
of CAPES and funding from the IIP. 
\end{acknowledgments}

\appendix

\section{Feynman Diagram in the spin susceptibility \label{feynman} }

\begin{figure}[h]
\includegraphics[width=8cm]{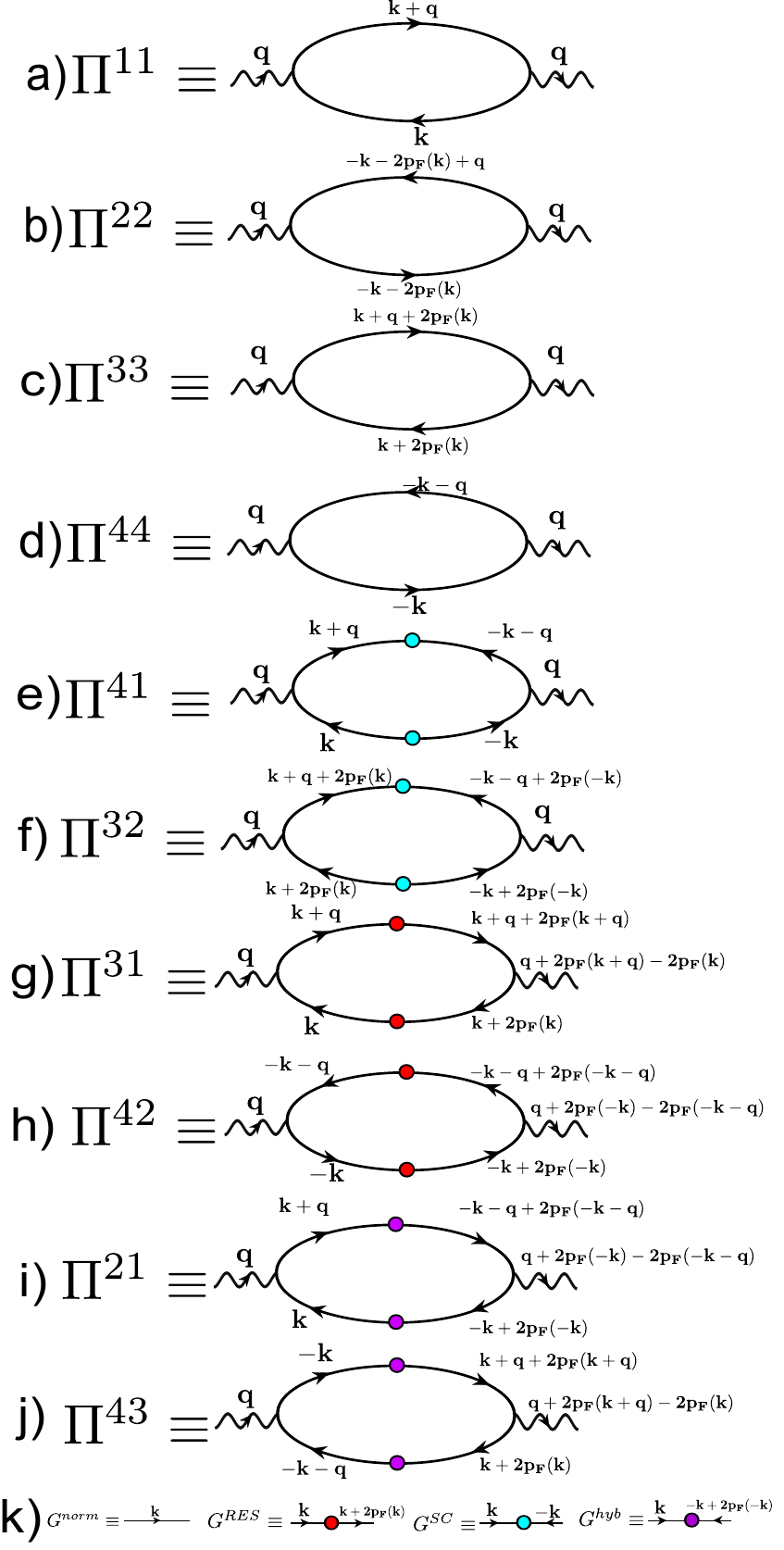} \caption{\label{Giagram} Polarizations $\Pi^{ij}$ that contribute to the
bare polarization $\chi_{0}$ (see equation (\ref{eq:suspol})). In
a) to d) are the diagrammatic representation of the polarization with
normal contribution. In e) an f) are presented the contribution of
superconducting state. In g) and h) are presented the contribution
of the RES. In i) and j) are shown the mixed SC-RES contribution.
The contribution of the RES and SC-RES mixed polarization (diagrams
from g) to j)) only exist for $q$ close to $\left(0,0\right)$ and
$\left(\pi,\pi\right)$. In k) are presented the diagrammatic representation
of the Green function.}
\end{figure}

We consider the spin susceptibility originating the t-J model \cite{Brinckmann99,Brinckmann01}
which writes: 
\begin{equation}
\chi_{S}(\omega,\mathbf{q})=\frac{\chi_{S}^{0}(\omega,\mathbf{q})}{1+J(\mathbf{q})\chi_{S}^{0}(\omega,\mathbf{q})}\label{eq:susa}
\end{equation}
with $J(\mathbf{q})=J_{0}\left(\text{cos}(q_{x}a_{0})+\text{cos}(q_{y}a_{0})\right)$.
In the equation (\ref{eq:sus}), $\chi_{S}^{0}$ is the bare polarization
bubble constructed from the Green's function and $J(\mathbf{q})$
is super-exchange interaction. Note that momentum dependence of the
super-exchange term $J(\mathbf{q})$ originates the exchange between
near neighbor Copper site. The bare polarization can be evaluated
by the formula \cite{Norman07,Schrieffer64}: 
\begin{equation}
\chi_{S}^{0}\left(\omega,\mathbf{q}\right)=-\frac{T}{2}\sum_{\epsilon,\mathbf{k}}\text{Tr}\left[\hat{G}\left(\omega+\varepsilon,\mathbf{k+q}\right)\hat{G}\left(\varepsilon,\mathbf{k}\right)\right]\label{eq:susgena}
\end{equation}
where $\varepsilon(\omega)$ is the fermionic (bosonic) Matsubara
frequency, $\mathbf{k(q)}$ is the impulsion, $T$ the temperature
and Tr means Trace of the Green function matrix $\hat{G}$. The relation
(\ref{eq:susgena}) describes the whole polarization of the system
that is the sum of the polarizations $\Pi$: 
\begin{equation}
\chi_{S}^{0}=\frac{1}{8}\left(\sum_{i,j}\Pi^{ij}\right)\label{eq:suspol}
\end{equation}
where $\Pi^{ij}$ are the polarizations described by the diagrams
of the Fig.\ref{Giagram} with $\Pi^{ij}=-T\sum_{\varepsilon,{\bf {k}}}\left[G^{ij}\left(\omega+\varepsilon,\mathbf{k+q}\right)G^{ij}\left(\varepsilon,\mathbf{k}\right)\right]$
with $\Pi^{ij}=\Pi^{ji}$ for $j\neq i$. $\Pi^{11(44)}$ (diagram
(a) (and (d)) of Fig.\ref{Giagram}) is the response of the electrons
(holes) with momentum ${\bf {k}}$ while $\Pi^{22(33)}$ (diagram
(b) (and (c)) of Fig.\ref{Giagram}) is the response of the electrons
(holes) with momentum ${\bf {k}+\pff}$. The polarization $\Pi^{41(32)}$
(diagram (e) (and (f)) of Fig.\ref{Giagram}) is the response of the
Cooper pairs while $\Pi^{31(42)}$ (diagram (g) (and (h)) of Fig.\ref{Giagram})
is the response of the particle-hole pairs. The polarization $\Pi^{21(43)}$
(diagram (i) (and (j)) of Fig.\ref{Giagram}) is the mixed SC-RES
response. Note that the superconducting coherent factors comes from
the terms $\Pi_{SC}^{41(32)}$.

As shown in the diagram (g) to (j) of Fig.\ref{Giagram}, the outgoing
external vector depends on the difference $\bar{\delta}_{\pf}=\pffq-\pff$.
In order to these diagrams to contribute to the global polarization
($\Pi^{21(31,42,43)}\neq0$), this difference must vanish, $\bar{\delta}_{\pf}=\mathbf{0}$.
Obviously, this difference vanishes for ${\bf {q}=0}$. This difference
also vanishes for ${\bf {q=Q}}$. The RES polarization contributes
around ${\bf {q}\approx0}$ and ${\bf {q\approx Q}}$ but will vanish
if ${\bf {q}}$ is far from $\bf{0}$ or ${\bf {Q}}$. To take modelize
this effect, we introduce a momentum dependent function in the relation
(\ref{eq:suspol}) which transforms itself as : 
\begin{align}
\chi_{S}^{0}=\frac{1}{8}\left(\Pi^{11}+\Pi^{22}+\Pi^{33}+\Pi^{44}+2(\Pi^{32}+\Pi^{41})\right.\nonumber \\
\left.+2f(\mathbf{q})(\Pi_{RES}^{21}+\Pi_{RES}^{31}+\Pi_{RES}^{42}+\Pi_{RES}^{43})\right)\label{polPGgen}
\end{align}
where $f(\mathbf{q})$ acts on the RES and SC-RES mixed polarizations. The function $f(\mathbf{q})$
has the form : 
\begin{equation}
f(\bq)=\frac{1}{1+V(sin^{2}(q_{x}a)+sin^{2}(q_{y}a))}\label{Lor}
\end{equation}
which is a Lorentzian centered in ${\mathbf {q}=(0,0)}$ and ${\bf {q}={\bf {Q}=(\pi,\pi)}}$
whose width can be tuned by the parameter $V$. If $V$ tends toward
zero, the function $f({\bf {q})}$ uniformly tends to unity. If $V$
tends toward infinity, the function $f({\bf {q})}$ is a dirac distribution
centered in $(0,0)$ and $(\pi,\pi)$. The effect of the function
$f$ on the spin susceptibility is detailed in the Appendix \ref{feffect}.

\section{Effect of the $f$ function on the spin susceptibility $\chi_{S}$
around ${\bf {Q}}$ \label{feffect}}

\begin{figure}[h]
a)\includegraphics[scale=0.65]{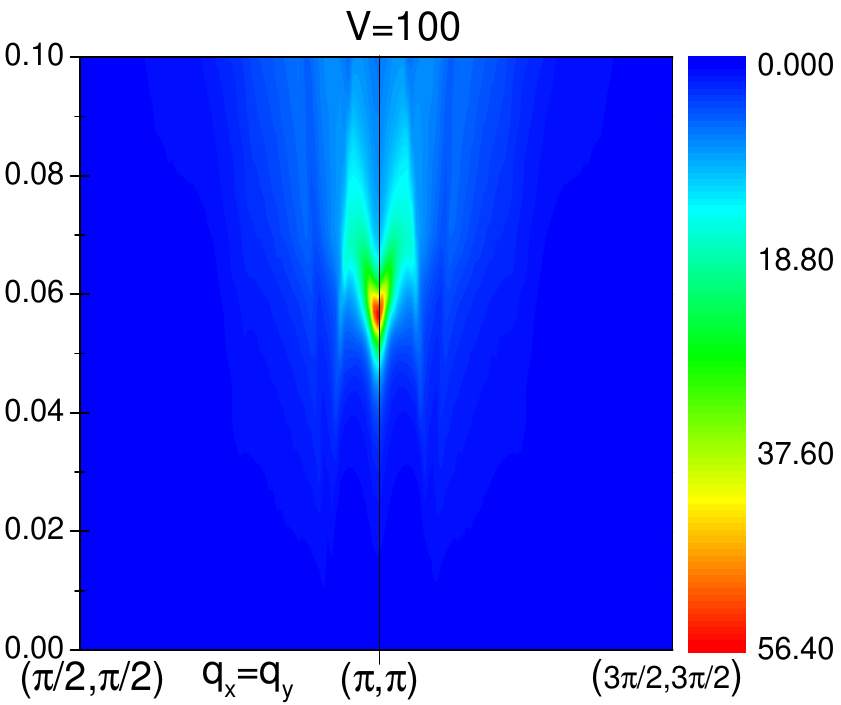} 
b)\includegraphics[scale=0.65]{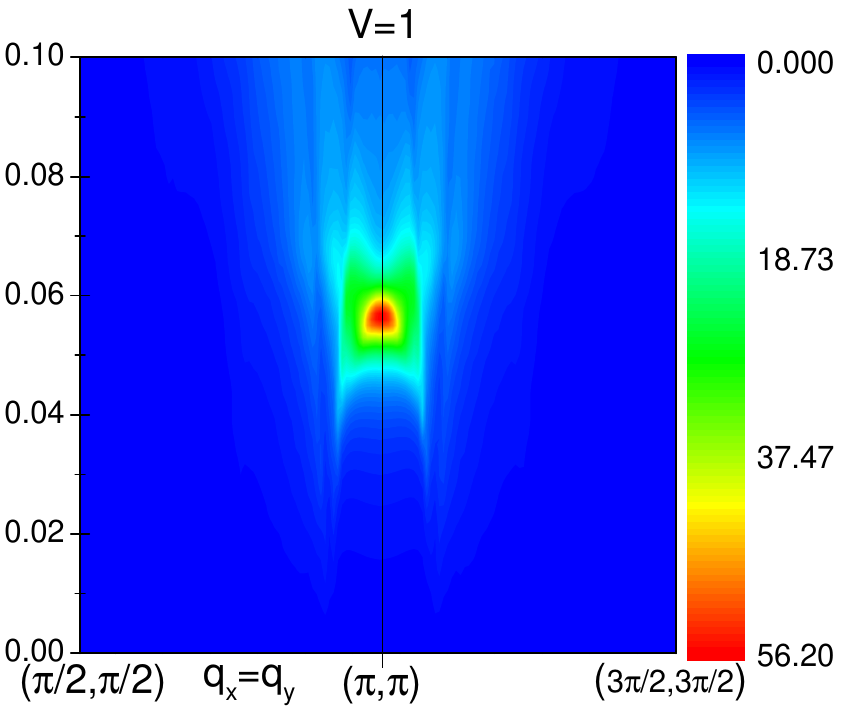} 
c)\includegraphics[scale=0.65]{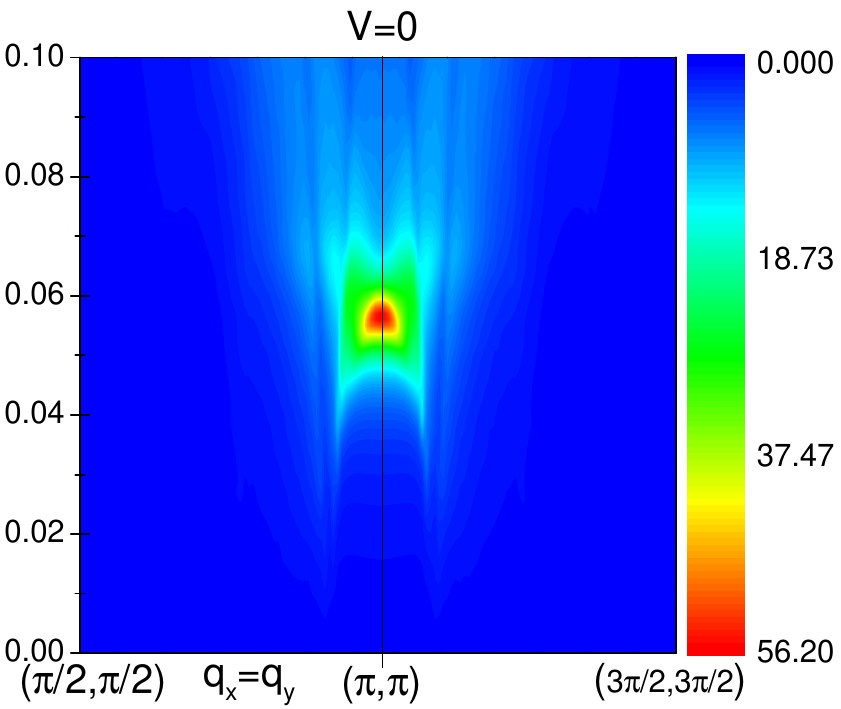} \caption{\label{effectV} Amplitude of the Imaginary part of the spin susceptibility
$\chi_{S}$ as a function of $\omega$, for $q_{y}=q_{x}$ and $q_{x}$
from $-\pi/2a$ to $3\pi/2a$ at $p=0.16$ for $J_{0}=0.2225$ as
a function of the parameter V. In a) V=100, in b) V=1 and in c) V=0.
The solid line is set at $q=(\pi/a,\pi/a)$. The susceptibility $\chi_{S}$
is calculated in the RES state at $p=0.16$ and the order parameter
magnitude are $\Delta_{SC}^{0}=0meV$ and $\Delta_{RES}^{0}=100meV$.
The value of V affects the width of the function $f$. For $V=0$,
we clearly observe two branches from either side of momentum $\bbq$
which are cut with higher value of $V$.}
\end{figure}

In this section we present the effect of the width of the function
$f$ on the spin susceptibility $\chi_{S}$. The function $f$ is
a Lorentzian whose width can be tune by the value of the parameter
$V$ (see formula \ref{Lor}). If $V$ vanishes then
$f$ is uniformly unity, $f=1$. If $V$ tends toward infinity then
$f$ becomes a Dirac function centered in $\bbq$. In the Fig. \ref{effectV},
we present the spin susceptibility $\chi_{S}$ as a function of the
parameter $V$. We observe that the for $V=0$ (Fig. \ref{effectV}
c)), the energy fluctuation in the RES looks like the one in the pure
SC state \cite{Norman07} with the two branches from either side of
the momentum $\bbq$ but with a particle-hole continuum at $\bbq$.
When the parameter $V$ increases (Fig. \ref{effectV} a) and b)),
the two branches are completely lowered and only the resonance at
$\bbq$ remains. 

\section{Effect of the SC order parameter momentum dependence on the spin susceptibility}
\label{gapeffect}
\begin{figure}[h]
a)\includegraphics[scale=0.85]{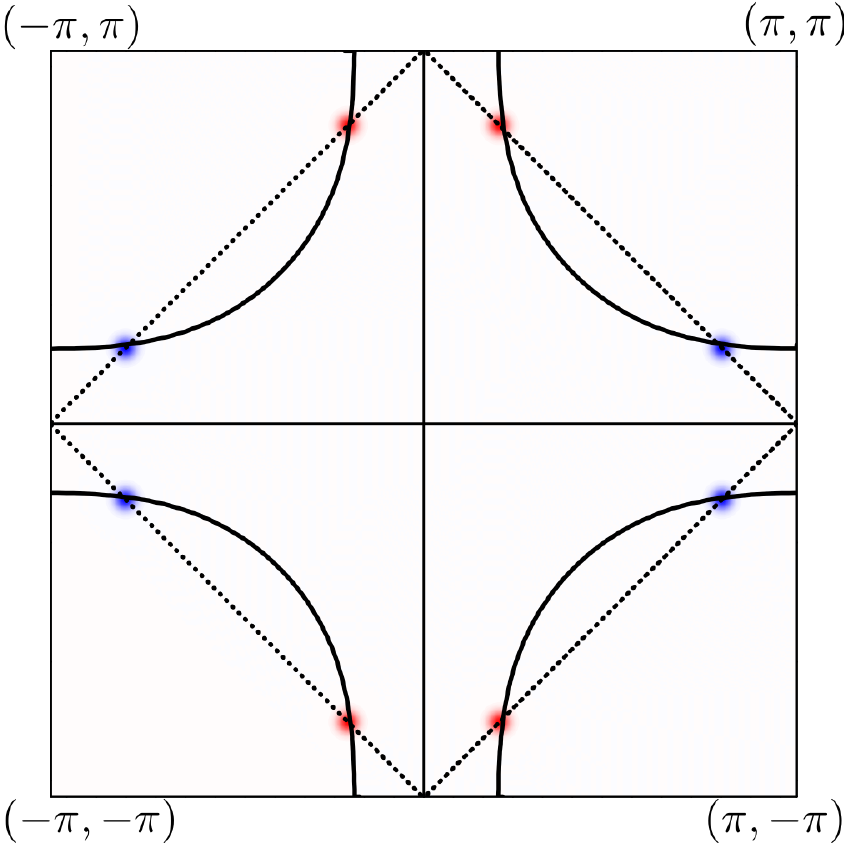} 
b)\includegraphics[scale=0.85]{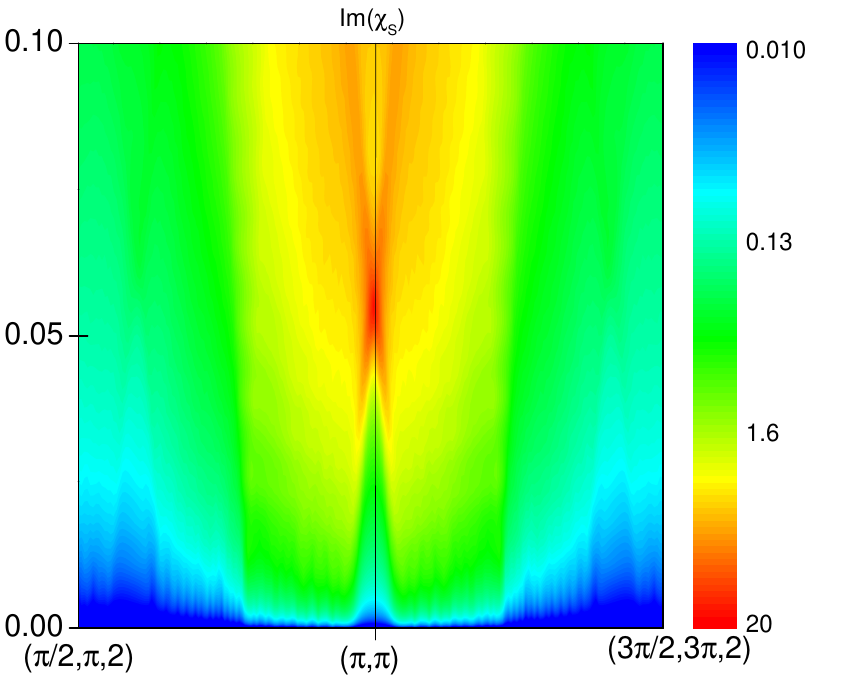} 
\caption{\label{DepSCgap} a) Amplitude of the superconducting gap in the first Brillouin zone. The superconducting gap is centered on the hot-spots. b) Amplitude of the Imaginary part of the spin susceptibility $\chi_{S}$ as a function of $\omega$, for $q_{y}=q_{x}$ and $q_{x}$ from $-\pi/2a$ to $3\pi/2a$ at $p=0.16$ for $J_{0}=0.2225$.  The amplitude is centered in the $(\pi,\pi)$ and the X-shape disappears.}
\end{figure}

In this section, we present the effect of the momentum dependence of the SC order parameter on the form of the spin susceptibility. If we consider a SC gap centered only on the hot-spot (see Fig. \ref{DepSCgap} a)), the spin susceptibility is maximal around the vector $\left(\pi,\pi\right)$ only and the X shape disappears, as shown in Fig. \ref{DepSCgap} b).

\bibliographystyle{apsrev4-1}
\bibliography{Cuprates}

\end{document}